\documentclass[12pt]{article}

\usepackage{graphicx,a4,epsfig,rotate,amsmath,amssymb,citesort}
\usepackage{latexsym,color,slashed}
\newcommand{\be}{\begin{equation}}
\newcommand{\ee}{\end{equation}}
\newcommand{\bea}{\begin{eqnarray}}
\newcommand{\eea}{\end{eqnarray}}
\newcommand{\eq}{\begin{eqnarray}}
\newcommand{\en}{\end{eqnarray}}

\newcommand{\ed}{\end{document}}
\newcommand{\bc}{\begin{center}}
\newcommand{\ec}{\end{center}}

\begin{document}

\thispagestyle{empty}

\begin{center}

\vspace{3cm}
{\Large{\bf Matrix elements of unstable states}}

\vspace{0.5cm}

\vspace{0.5cm}

V. Bernard$^a$,
D. Hoja$^b$,
U.-G. Mei{\ss}ner$^{b,c}$ and
A.~Rusetsky$^b$

\vspace{2em}

\begin{tabular}{c}
$^a\,${\it Institut de Physique Nucl\'eaire, CNRS/Univ. Paris-Sud 11 (UMR 8608),}\\
{\it F-91406 Orsay Cedex, France}\\[2mm]
$^b\,${\it Helmholtz-Institut f\"ur Strahlen- und Kernphysik (Theorie) and}\\
{\it  Bethe Center
for Theoretical Physics,  Universit\"at Bonn,}\\
{\it  D-53115 Bonn, Germany}\\[2mm]
$^c${\it Forschungszentrum J\"ulich, J\"ulich Center for Hadron Physics,}\\
{\it Institut f\"ur Kernphysik  (IKP-3) and 
Institute for Advanced Simulation (IAS-4),}\\
{\it D-52425 J\"ulich, Germany}\\
\end{tabular}

\end{center}

\vspace{1cm}

{\abstract
{
Using the language of  non-relativistic effective Lagrangians, we
formulate a systematic framework for the calculation of resonance
matrix elements in lattice QCD. The generalization of the L\"uscher-Lellouch
formula for these matrix elements is derived. We further discuss in detail
the procedure of the analytic continuation of the resonance matrix elements
into the complex energy plane and investigate the infinite-volume limit.
}}

\vskip1cm

{\footnotesize{\begin{tabular}{ll}
{\bf{Pacs:}}$\!\!\!\!$& 11.10.St, 11.15.Ha, 13.40.Gp\\
{\bf{Keywords:}}$\!\!\!\!$& Resonances in lattice QCD,
field theory in a finite volume,\\$\!\!\!\!$ &Non-relativistic
effective field theory, Form factors
\end{tabular}}
}
\clearpage


\section{Introduction}

The calculation of  matrix elements involving unstable states
has already been addressed in lattice QCD. As examples, we mention the recent
papers \cite{Gurtler:2008zz,Alexandrou:2011ga,Alexandrou:2010tj}, which
deal with the electromagnetic form factor of the $\rho$-meson,
as well as the electromagnetic and axial-vector form factors of the 
$\Delta$-resonance and the $N\Delta\gamma$ transition vertex. Electromagnetic
and axial transition form factors for the Roper resonance have also been
studied~\cite{em-Roper}. Moreover, 
we expect that the
number of such investigations will substantially grow in the nearest
future due to a growing interest in the study of the excited states.

Even if one argues that the quark (pion) masses in the above lattice simulations
are large, so that all resonances are in fact stable particles,
various conceptual questions  arise: 

\begin{itemize}
\item[i)]
It is clear that we are ultimately
interested in simulations carried out at the physical quark masses.
Is it possible (at least in principle) to tune the quark mass continuously
until it reaches the physical value?

\item[ii)]
In the continuum field theory, any matrix element with {\em resonance} states
is defined through an analytic continuation of the three-point
Green function into the complex plane
$P^2\to s_R$, where $P$ denotes the pertinent four-momentum
and  $s_R$ is the resonance pole position in the complex plane (its real
and imaginary parts are related to the mass and the width of a resonance).
What is the analog of this procedure in lattice field theory?

\item[iii)]
Once this procedure is defined, 
what is the volume dependence of the measured form factors?

\end{itemize}

In this paper, we address these questions in detail.
In order to formulate the problem in a more transparent manner, let
us first define what is meant by resonance matrix elements in the
continuum field theory and on the lattice. We start with the continuum field
theory and, for simplicity, concentrate on the scalar case.
Consider an arbitrary (local or non-local) scalar
operator $O(x)$ which has the internal quantum numbers
of a given resonance. The statement that a resonance is present is equivalent
to the claim that the two-point function
\eq
D(P^2)=i\int d^4x\,e^{iPx}\langle 0|TO(x)O^\dagger(0)|0\rangle
\en
has a pole in the complex variable $P^2$ on the lower half of the 
second Riemann sheet at 
$P^2=s_R$:
\eq\label{eq:twopoint-pole}
\lim_{P^2\to s_R}D(P^2)=\frac{B_R}{s_R-P^2}+\mbox{regular terms at }P^2\to s_R\, .
\en
The real and imaginary parts of $s_R$ are related to the resonance mass
$m_R$ and the width $\Gamma_R$, according to
$\mbox{Re}\,s_R=m_R^2-\Gamma_R^2/4$, $\mbox{Im}\,s_R=-m_R\Gamma_R$.

In order to define resonance matrix elements\footnote{The following 
discussion is a straightforward adaptation of the procedure which has
been used to define the matrix elements in the case of {\em stable} composite
objects, see, e.g., Refs.~\cite{Mandelstam,HuangWeldon}.}, say, of the
electromagnetic current $J_\mu$, we consider the following three-point
function: 
\eq
V_\mu(P,Q)=i^2\int d^4x\,d^4y\,e^{iPx-iQy}\langle 0|TO(x)J_\mu(0)O^\dagger(y)|0\rangle\, .
\en
The form factor of a resonance is then {\em defined} as
\eq\label{eq:form}
i(P+Q)_\mu F((P-Q)^2)\doteq\langle P|J_\mu(0)|Q\rangle
=\lim_{P^2,Q^2\to s_R} B_R^{-1/2} (s_R-P^2)V_\mu(P,Q)(s_R-Q^2) B_R^{-1/2}\, ,
\en
where $B_R$ is the residue at the 
resonance pole, see Eq.~(\ref{eq:twopoint-pole}).
Note that the matrix element displayed in Eq.~(\ref{eq:form}) should be 
understood as a mere notation: in the spectrum, there exists no isolated
resonance state with a definite momentum. Moreover, as it is clear from 
Eq.~(\ref{eq:form}), this definition of the resonance matrix elements
necessarily implies an analytic continuation into the complex plane.
We would like to stress that  we are not aware of any consistent field-theoretical 
prescription, where the analytic continuation would not be employed.

Let us now ask the question, how such resonance matrix elements could be
evaluated on the lattice (at least, in principle). As it is well known,
a resonance does not appear as an isolated energy level. There exist
alternative approaches to the problem of extracting resonance characteristics
(the mass and the width)
from the measured quantities on the Euclidean lattice. In this paper,
we work within L\"uscher's finite-volume framework~\cite{Luescher-torus}\footnote{
 At present, 
L\"uscher's approach~\cite{Luescher-torus} has been widely used to obtain
scattering phase shifts from the  energy spectrum in a finite 
volume. The resonance position can be then established by using the measured
phase shift. The procedure can be directly generalized to the case of
multi-channel scattering~\cite{He,lage-KN,lage-scalar}. Moreover, in Ref.~\cite{oset}
it has been argued that the use of the physical input based on
unitarized Chiral Perturbation Theory may facilitate the extraction of the
resonance poles from the lattice data (the method has been subsequently applied
to different physical problems in Refs.~\cite{oset-others,kappa}).
Recently, a generalization of L\"uscher's approach in the presence of 
3-particle intermediate states has been proposed~\cite{Polejaeva-2}.
Other approaches to the determination of resonance pole positions
imply the study of the two-point function at finite 
times~\cite{Michael,polejaeva}, as well as reconstructing the spectral
density by using the maximal entropy method~\cite{entropy}. 
The application of different approaches to the extraction of the resonance 
properties from the lattice data has been carried out recently in 
Ref.~\cite{Peardon}. Last but not least, the 
finite volume approach has been applied to study
the two-particle decay matrix elements on the lattice~\cite{Lellouch,Sachrajda,Meyer},
including the case of multiple channels~\cite{Sharpe}.}.
In order to calculate the matrix element on the lattice, one usually considers
the following three-point function
\eq
\tilde V_\mu({\bf P},t';{\bf Q},t)=\langle 0|TO_{\bf P}(t')J_\mu(0)
O_{\bf Q}^\dagger(t)|0\rangle\, ,
\en
where
\eq
O_{\bf P}(t')=\sum_{\bf x}e^{-i{\bf Px}}O({\bf x},t')\, ,\quad\quad
O_{\bf Q}^\dagger(t)=\sum_{\bf y}e^{i{\bf Qy}}O^\dagger({\bf y},t)\, .
\en
In addition, we define:
\eq
D({\bf P},t)=\langle 0|TO_{\bf P}(t)O^\dagger_{\bf P}(0)|0\rangle\, .
\en
The matrix element of the electromagnetic current between the {\em ground-state
vectors} in a channel with the quantum numbers of the operator $O(x)$,
moving with the 3-momenta ${\bf P}$ and ${\bf Q}$, respectively,
is given by
\eq\label{eq:E0}
\langle E_0({\bf P})|J_\mu(0)|E_0({\bf Q})\rangle
=\displaystyle \lim_{\substack{t'\to+\infty\\ t\to-\infty}}\tilde V_\mu({\bf P},t';{\bf Q},t)
\sqrt{\frac{D({\bf Q},t')D({\bf P},t)}
{D({\bf Q},t)D({\bf Q},t'-t)D({\bf P},t-t')D({\bf P},t')}} .
\en
Using the generalized eigenvalue equation method, 
the matrix elements between the excited state vectors $|E_n({\bf P})\rangle$
 can be also defined in a similar manner.

If the ground state of a system corresponds to a stable particle, then
Eq.~(\ref{eq:E0}) indeed yields the form factor
of a stable particle in the infinite-volume limit, which in this case
is well defined. However, the situation in case of resonances is 
conceptually different. The easiest way to see this is to note that
in the infinite-volume limit the energy of any state $|E_n({\bf P})\rangle$ 
tends to the two-particle threshold energy. 
In other words, any given energy level decays
into the free particle levels
in the limit $L\to\infty$ (here, $L$ denotes the size of a spatial box). 
Moreover, as shown in Ref.~\cite{1+1} (in case
of $1+1$ dimensions), the matrix elements measured for any given level follow
a similar pattern. For example, the magnetic moment tends to the sum of
the magnetic moments of the free particles in the limit $L\to\infty$.
Obviously, this is not the result that we wish to extract from lattice data.

As mentioned above, using L\"uscher's approach, it is possible to determine the
resonance pole position in the complex plane from the measured finite-volume
(real) energy spectrum. This position stays put (up to exponentially 
suppressed corrections in $L$) in the limit $L\to\infty$, despite the fact
that all individual levels collapse towards threshold in this limit.
The aim of the present paper is to formulate a similar approach for the
matrix elements, and to ensure that the matrix elements that are extracted
with the help of such a procedure coincide with the infinite-volume matrix 
elements, e.g., given in Eq.~(\ref{eq:form}), up to  exponentially suppressed
corrections. 

The goal, stated above, will be achieved by a systematic use of  
non-relativistic effective field theory (EFT) in a finite volume. In particular,
we shall calculate the quantity in Eq.~(\ref{eq:E0}), which can be measured
on the lattice, within  non-relativistic EFT, and shall identify a piece
in this expression, whose infinite-volume limit coincides with the resonance
matrix element in the infinite volume we are looking for.

The paper is organized as follows: In section~\ref{sec:moving}
we formulate a covariant non-relativistic EFT in a moving frame and
re-derive the Gottlieb-Rummukainen~\cite{gottlieb} formula within this approach.
The extraction of a resonance pole position is discussed in detail.
In section~\ref{sec:LL} we give a short re-derivation of
the L\"uscher-Lellouch formula~\cite{Lellouch}, as another application of the non-relativistic
EFT methods. Further,
in section~\ref{sec:vertex} we evaluate the vertex function in the 
non-relativistic EFT. The infinite-volume limit of different terms in the expression
of the vertex function is analyzed in detail in section~\ref{sec:fixed}, where
particular attention is paid to the so-called fixed singularities that
emerge in a result of analytic continuation of L\"uscher's zeta-function 
into the complex plane in $3+1$ dimensions. 
The prescription for calculating the 
resonance matrix elements is given in section~\ref{sec:subtract}.
Section~\ref{sec:concl} contains our conclusions.

\section{Extraction of the resonance poles in moving frames}
\label{sec:moving}

The initial and final states in a form factor have  non-zero momenta. For this
reason, one has to formulate a procedure for extracting resonance pole 
positions in  moving frames. Within  potential quantum mechanics, this
has been done in Refs.~\cite{gottlieb}, see also Ref.~\cite{Fu,Prelovsek}
for the generalization to the non-equal mass case.
Refs.~\cite{Sachrajda,Davoudi} address the same problem in a
field-theoretical setting. Finally, in Ref.~\cite{Bob}, a full group-theoretical
analysis of the resulting equation has been performed, including the case of
particles with spin. Below, we shall briefly
re-derive this result within the non-relativistic EFT along the lines similar
to Refs.~\cite{Beane,lage-distributions}, where the treatment was restricted
to the rest frame. At a later stage, the same approach will be used for the
calculation of the matrix elements.

In the treatment of the moving frames it is very convenient to use the
covariant form of the non-relativistic EFT which has been introduced in 
Ref.~\cite{cusp1} and was discussed in detail in Ref.~\cite{cuspbig}. Assume,
for simplicity, that we deal with two elementary scalar fields $\Phi_{1,2}$
with  masses $m_{1,2}$, respectively. 
The Lagrangian is given in the following form:
\eq\label{eq:lag}
 {\cal L}&=&\sum_{i=1,2}\Phi_i^\dagger2W_i(i\partial_t-W_i)\Phi_i
+C_0\Phi_1^\dagger\Phi_1\Phi_2^\dagger\Phi_2
\nonumber\\[2mm]
&+&C_1\bigl((\Phi_1^\dagger)^\mu(\Phi_2^\dagger)_\mu\Phi_1\Phi_2-m_1m_2\Phi_1^\dagger\Phi_1\Phi_2^\dagger\Phi_2
+\mbox{h.c.}\bigr)
\nonumber\\[2mm]
&+&C_2\bigl(\Phi_1^\dagger(\Phi_2^\dagger)^\mu-(\Phi_1^\dagger)^\mu\Phi_2^\dagger)
((\Phi_1)_\mu\Phi_2-\Phi_1(\Phi_2)_\mu)
+\cdots\, ,
\en
where $\Phi_i,~i=1,2$ denote the non-relativistic field operators,
$W_i=\sqrt{m_i^2+\triangle}$ 
are the energies of the particles 
(here, $\triangle \doteq \nabla^2$),
and
\eq\label{eq:calP}
(\Phi_i)_\mu=({\cal P}_i)_\mu\Phi_i\, ,\quad
(\Phi_i^\dagger)_\mu=({\cal P}_i^\dagger)_\mu\Phi_i^\dagger\, ,\quad\quad
({\cal P}_i)_\mu=(W_i,-i\nabla)\, ,\quad
({\cal P}_i^\dagger)_\mu=(W_i,i\nabla)\, .
\en
Further, the
ellipses stand for terms containing at least four space derivatives. 
To set up the power-counting rules we introduce, as in Refs.~\cite{cusp1,cuspbig},
a generic small parameter $\epsilon$ and count each 3-momentum
as ${\bf p}_i=O(\epsilon)$, whereas the masses are counted as $m_i=O(1)$. 
The Lagrangian given in Eq.~(\ref{eq:lag}) contains all allowed 
{\em explicitly Lorentz-invariant} terms\footnote{Note that in the 
conventional non-relativistic theory the number of the allowed terms 
at a given order
in $\epsilon$ is much larger, because these terms are not restricted
by the requirement of Lorentz-invariance. 
At the end, however, matching to the
relativistic amplitude should be performed that effectively imposes such
constraints on the low-energy couplings, because the number of physically 
independent low-energy parameters in the relativistic amplitude is smaller.
In this way, the constraints are imposed in a perturbative manner, order
by order in $\epsilon$. On the contrary, in our approach, we impose the
requirement of the Lorentz invariance from the beginning and avoid the
introduction of the constraints at all. The key property which
allows us to do this is that in our approach (unlike the conventional 
framework) non-relativistic loops are Lorentz-invariant by itself, 
so it suffices 
to impose Lorentz-invariance at tree level only. For more details,
we refer the reader to Ref.~\cite{cuspbig}. The method of matching to the
relativistic theory was already used in the construction of the
heavy-baryon chiral effective Lagrangian in Ref.~\cite{Bernard:1992qa}.} 
up-to-and-including
$O(\epsilon^2)$, and the omitted terms are of order $\epsilon^4$.

The non-relativistic couplings $C_0,C_1,C_2,\cdots$, which are 
present in the Lagrangian,  are directly related to
the effective-range expansion parameters for  $1+2\to 1+2$ 
elastic scattering (scattering length, effective range, etc), 
see Refs.~\cite{cusp1,cuspbig}.  We would like to remind the reader
here that the theory described by the Lagrangian given in Eq.~(\ref{eq:lag})
conserves particle number, so it can be applied in the elastic region only.

The Feynman rules, which are produced by the Lagrangian~(\ref{eq:lag}), 
should be amended by a prescription which states that the integrand
in each Feynman integral is expanded in 3-momenta, each term is integrated
by using  dimensional regularization and the result is summed up 
again~\cite{cusp1,cuspbig}. Below, we shall consider the theory in a finite
volume. It is easy to see that, for consistency, one should apply the
same prescription, replacing the dimensionally regularized integrals by
sums over  discrete momenta. In particular, one has
to discard everywhere discrete sums over  polynomials in momenta, 
in accordance with the similar infinite-volume prescription in the dimensionally
regularized theory.

Let us start in the infinite volume.
Using the above Feynman rules, it is straightforward to ensure that the
scattering $T$-matrix in the infinite volume in an arbitrary moving frame
obeys the Lippmann-Schwinger (LS) equation:
\eq\label{eq:LS}
&&T({\bf p}_1,{\bf p}_2;{\bf q}_1,{\bf q}_2)=
-V({\bf p}_1,{\bf p}_2;{\bf q}_1,{\bf q}_2)-
\int\frac{d^d{\bf k}_1}{(2\pi)^d2w_1({\bf k}_1)}
\frac{d^d{\bf k}_2}{(2\pi)^d2w_2({\bf k}_2)}\,
\nonumber\\[2mm]
&\times& (2\pi)^d\delta^d({\bf p}_1+{\bf p}_2-{\bf k}_1-{\bf k}_2)\,
\frac{V({\bf p}_1,{\bf p}_2;{\bf k}_1,{\bf k}_2)
T({\bf k}_1,{\bf k}_2;{\bf q}_1,{\bf q}_2)}{w_1({\bf k}_1)+w_2({\bf k}_2)
-w_1({\bf p}_1)-w_2({\bf p}_2)-i0}\, ,
\en
where $w_i({\bf l})=\sqrt{m_i^2+{\bf l}^2}$ and the potential is given
by the matrix element of the interaction Hamiltonian, which is derived from
the Lagrangian (\ref{eq:lag}) by the canonical procedure, between the
two-particle states
\eq
\langle {\bf p}_1,{\bf p}_2|H_{\rm I}|{\bf q}_1,{\bf q}_2\rangle
&=& (2\pi)^3\delta^3({\bf p}_1+{\bf p}_2-{\bf q}_1-{\bf q}_2)
V({\bf p}_1,{\bf p}_2;{\bf q}_1,{\bf q}_2)\, .
\en
Note that we have used dimensional regularization in Eq.~(\ref{eq:LS}).
The parameter $d$ denotes the number of space dimensions (at the end of
calculations, $d\to 3$).

By construction, the potential $V$ is a Lorentz-invariant low-energy
polynomial that depends
only on scalar products of the 4-momenta. The first few terms
in the expansion are given by
\eq\label{eq:VC0}
-V({\bf p}_1,{\bf p}_2;{\bf q}_1,{\bf q}_2)
=C_0+C_1(p_1p_2+q_1q_2-2m_1m_2)+C_2(p_2-p_1)(q_2-q_1)+O(\epsilon^4)\, ,
\en
where, e.g., $p_1p_2=w_1({\bf p}_1)w_2({\bf p}_2)-{\bf p}_1{\bf p}_2$, etc. 
In general, defining the center-of-mass (CM) and relative momenta, according to
\eq\label{eq:CM}
\!\!&&\!\!\!\!
P=p_1+p_2\, ,\quad p=\mu_2p_1-\mu_1p_2\, ,\quad\quad
\mu_{1,2}=\frac{1}{2}\biggl(1\pm\frac{m_1^2-m_2^2}{P^2}\biggr)\, ,\quad
p^2=\frac{\lambda(P^2,m_1^2,m_2^2)}{4P^2}\, ,
\nonumber\\[2mm]
\!\!&&\!\!\!\!
Q=q_1+q_2\, ,\quad q=\mu_2'q_1-\mu_1'q_2\, ,\quad\quad
\mu'_{1,2}=\frac{1}{2}\biggl(1\pm\frac{m_1^2-m_2^2}{Q^2}\biggr)\, ,\quad
q^2=\frac{\lambda(Q^2,m_1^2,m_2^2)}{4Q^2}\, ,
\en
where $\lambda(x,y,z)$ denotes the K\"all\'en triangle function, 
it can be seen that $V$ is a low-energy
polynomial of six independent Lorentz-invariant arguments
$p^2$, $q^2$, $pq$, $PQ$, $Pq$, $pQ$. The original arguments 
$p_1p_2$, $p_1q_1$, $p_1q_2$, $p_2q_1$, $p_2q_2$, $q_1q_2$ can 
be expressed through  linear
combinations of these arguments with  coefficients, which themselves are
low-energy polynomials.

Consider now the partial-wave expansion of the potential. To this end,
we define the momenta boosted to the CM frame (note that the boost
velocity is different in the initial and the final states, because
the potential is generally off the energy shell):
\eq
&&
{\bf p}^*={\bf p}+{\bf P}\biggl((\gamma-1)\frac{{\bf p}{\bf P}}{{\bf P}^2}
-\gamma v \frac{p_0}{|{\bf P}|}\biggr)\, ,\quad
p_0^*=\gamma p_0-\gamma v\frac{{\bf p}{\bf P}}{|{\bf P}|}=0\, ,
\nonumber\\[2mm]
&&
P^*_\mu=(\sqrt{P^2},{\bf 0})\, ,\quad\quad 
v=\frac{|{\bf P}|}{P_0}\, ,\quad\gamma=(1-v^2)^{-1/2}\, ,
\nonumber\\[2mm]
&&
{\bf q}^*={\bf q}+{\bf Q}\biggl((\gamma'-1)\frac{{\bf q}{\bf Q}}{{\bf Q}^2}
-\gamma' v' \frac{q_0}{|{\bf Q}|}\biggr)\, ,\quad
q_0^*=\gamma' q_0-\gamma' v'\frac{{\bf q}{\bf Q}}{|{\bf Q}|}=0\, ,
\nonumber\\[2mm]
&&
Q^*_\mu=(\sqrt{Q^2},{\bf 0})\, ,\quad\quad
v'=\frac{|{\bf Q}|}{Q_0}\, ,\quad\gamma'=(1-(v')^2)^{-1/2}\, .
\en
Taking into account the fact that ${\bf P}={\bf Q}$ in the ``lab frame,''
it is straightforward to show that
\eq
pq&=&p^*q^*+O((P^0-Q^0)^2)=-{\bf p}^*{\bf q}^*+O((P^0-Q^0)^2)\, ,
\nonumber\\[2mm]
PQ&=&=P^*Q^*+O((P_0-Q_0)^2)=\sqrt{P^2}\sqrt{Q^2}+O((P_0-Q_0)^2)\, ,
\nonumber\\[2mm]
Pq&=&P^*q^*+O(P_0-Q_0)=0+O(P_0-Q_0)\, ,
\nonumber\\[2mm]
pQ&=&p^*Q^*+O(P_0-Q_0)=0+O(P_0-Q_0)\, .
\en
In addition, $p^2$ and $q^2$ can be expressed in terms of $P^2$ and $Q^2$, respectively.
This means that,
up to terms that vanish as $P_0\to Q_0$, the 
potential can be rewritten in the following form
\eq\label{eq:pw-potential}
-V({\bf p}_1,{\bf p}_2;{\bf q}_1,{\bf q}_2)
=-4\pi\sum_{lm}v_l(|{\bf p}^*|,|{\bf q}^*|)
{\cal Y}_{lm}({\bf p}^*){\cal Y}^*_{lm}({\bf q}^*)+O(P_0-Q_0)\, .
\en
Here, the function $v_l$ can be chosen to be 
real and symmetric with respect to its arguments,
i.e., Eq.~(\ref{eq:pw-potential}) describes a Hermitean potential. 
The quantity ${\cal Y}_{lm}({\bf p})$ is defined as
${\cal Y}_{lm}({\bf p})=|{\bf p}|^lY_{lm}(\hat {\bf p})$, where
$Y_{lm}$ are the usual spherical harmonics. The terms
that vanish as $P_0\to Q_0$ can be omitted from now on. The 
justification for this is the fact that the parameters in the potential
are determined by matching to the physical $S$-matrix elements (on shell),
order by order in the low-energy expansion. The omitted terms do not contribute
either at tree level or in loops (the latter because the {\em regular}
momentum integrals vanish in  dimensional regularization). Consequently,
one may consistently set these terms equal to zero from the beginning.

Performing now the partial-wave expansion in the amplitude
\eq\label{eq:pw-amplitude}
T({\bf p}_1,{\bf p}_2;{\bf q}_1,{\bf q}_2)
=4\pi\sum_{lm}t_l(|{\bf p}^*|,|{\bf q}^*|)
{\cal Y}_{lm}({\bf p}^*){\cal Y}^*_{lm}({\bf q}^*)\, ,
\en
substituting this expansion into the LS equation~(\ref{eq:LS}), 
and using the properties of dimensional regularization,
on the
energy shell $|{\bf p}^*|=|{\bf q}^*|=\lambda^{1/2}(s,m_1^2,m_2^2)/(2\sqrt{s})$ we get
\eq
t_l(s)=-v_l(s)-v_l(s)|{\bf p}^*|^{2l}G(s)t_l(s)\, ,
\en
where the obvious shorthand notations for the on-shell quantities
$v_l(s)=v_l(|{\bf p}^*|,|{\bf p}^*|)$ and
$t_l(s)=t_l(|{\bf p}^*|,|{\bf p}^*|)$ are used. The quantity $G(s)$ is given 
by~\cite{cusp1,cuspbig}:
\eq
G(s)=
\int\frac{d^d{\bf k}_1}{(2\pi)^d2w_1({\bf k}_1)}
\frac{d^d{\bf k}_2}{(2\pi)^d2w_2({\bf k}_2)}\,
\frac{(2\pi)^d\delta^d({\bf P}-{\bf k}_1-{\bf k}_2)}
{w_1({\bf k}_1)+w_2({\bf k}_2)-P_0-i0}
=\frac{i|{\bf p}^*|}{8\pi\sqrt{s}}\, .
\en
Further, unitarity gives:
\eq\label{eq:unitarity}
t_l(s)=\frac{8\pi\sqrt{s}}{|{\bf p}^*|^{2l+1}}\,e^{i\delta_l(s)}\sin\delta_l(s)\, ,
\quad\quad
v_l(s)=-\frac{8\pi\sqrt{s}}{|{\bf p}^*|^{2l+1}}\,\tan\delta_l(s)\, ,
\en
where $\delta_l(s)$ is the scattering phase.

The transition to the finite volume is performed in the ``lab frame''.
The momenta are discretized according to
\eq
{\bf k}_i=\frac{2\pi}{L}\, {\bf n}_i\, ,\quad\quad {\bf n}_i\in\mathbb{Z}^3\, .
\en
The partial-wave expansion of the potential does not change. However, since
the introduction of a cubic box breaks rotational symmetry, the partial-wave
expansion of the scattering amplitude has to be modified:
\eq
\label{eq:pw-amplitude-box}
T({\bf p}_1,{\bf p}_2;{\bf q}_1,{\bf q}_2)
=(4\pi)\sum_{lm,l'm'}t_{lm,l'm'}(|{\bf p}^*|,|{\bf q}^*|;{\bf P})
{\cal Y}_{lm}({\bf p}^*){\cal Y}^*_{l'm'}({\bf q}^*)\, .
\en
Substituting this expression into the Lippmann-Schwinger equation, 
on the energy shell we obtain:
\eq\label{eq:LS-L}
t_{lm,l'm'}(s;{\bf P})=-\delta_{lm,l'm'}v_l(s)
-4\pi\sum_{l''m''}v_l(s){\cal X}_{lm,l''m''}(s,{\bf P})t_{l''m'',l'm'}(s;{\bf P})\, ,
\en
where
\eq\label{eq:M-ini}
{\cal X}_{lm,l'm'}(s,{\bf P})=
\frac{1}{L^3}\sum_{{\bf k}_1}
\frac{{\cal Y}^*_{lm}({\bf k}^*){\cal Y}_{l'm'}({\bf k}^*)}
{2w_1({\bf k}_1)2w_2({\bf P}-{\bf k}_1)(w_1({\bf k}_1)+w_2({\bf P}-{\bf k}_1)-P_0)}\, .
\en
Next, we use the identity~\cite{cuspbig}
\eq
\frac{1}{4w_1w_2(w_1+w_2-P_0)}&=&
\frac{1}{2P_0}\frac{1}{{\bf k}^2-\dfrac{({\bf k}{\bf P})^2}{P_0^2}-({\bf p}^*)^2}
\nonumber\\[2mm]
&+&\frac{1}{4w_1w_2}\biggl(\frac{1}{w_1+w_2+P_0}-\frac{1}{w_1-w_2+P_0}
-\frac{1}{w_2-w_1+P_0}\biggr)\, ,
\en
where ${\bf k}={\bf k}_1+\mu_1{\bf P}$. One can straightforwardly check that
the term in the brackets does not become singular in the physical region. 
Using the regular summation theorem~\cite{luescher-2.},
one may then replace the sum over ${\bf k}_1$ in this term by the integral.
Further, to be consistent
with our prescription for the calculation of the Feynman integrals in
dimensional regularization, one should put these integrals to zero. 
After this, the expression for  ${\cal X}_{lm,l'm'}(s,{\bf P})$ takes the
following form:
\eq\label{eq:M-trans}
{\cal X}_{lm,l'm'}(s,{\bf P})=\frac{1}{2P_0}\,\frac{1}{L^3}\sum_{{\bf k}={\bf k}_1+\mu_1{\bf P}}
\frac{{\cal Y}^*_{lm}({\bf k}^*){\cal Y}_{l'm'}({\bf k}^*)}
{{\bf k}^2-\dfrac{({\bf k}{\bf P})^2}{P_0^2}-({\bf p}^*)^2}\, .
\en
In order to transform this equation further, let us define
the parallel and perpendicular components of the three vectors with respect
to the CM momentum ${\bf P}$. In particular, one may write
${\bf k}^*=(k^*_\parallel,{\bf k}^*_\perp)$, where 
$k^*_\parallel=(\gamma^*)^{-1}k_\parallel$, ${\bf k}^*_\perp={\bf k}_\perp$
and $\gamma^*=(1-(v^*)^2)^{-1/2}$, 
$v^*=|{\bf P}|/E^*=|{\bf P}|/(w_1({\bf k}^*)+w_2({\bf k}^*))$. 
Consequently, on the energy shell $E^*=P_0$ we obtain:
${\bf k}^*={\bf r}=(\gamma^{-1}k_\parallel,{\bf k}_\perp)$ with
$\gamma=(1-{\bf P}^2/P_0^2)^{-1/2}$. Up to  exponentially
suppressed terms, Eq.~(\ref{eq:M-trans}) now takes the form
\eq\label{eq:M}
&&{\cal X}_{lm,l'm'}(s,{\bf P})=\frac{(p^*)^{l+l'+1}}{32\pi^2\sqrt{s}}\,
i^{l-l'}\,{\cal M}_{lm,l'm'}(s,{\bf P})\, ,
\nonumber\\[2mm]
&&{\cal M}_{lm,l'm'}(s,{\bf P})=
\frac{(-)^l}{\pi^{3/2}\gamma}\,
\sum_{j=|l-l'|}^{l+l'}\sum_{s=-j}^j\frac{i^j}{\eta^{j+1}}\,
Z^{{\bf d}}_{js}(1;s)C_{lm,js,l'm'}\, ,
\nonumber\\[2mm]
&&C_{lm,js,l'm'}=(-)^{m'}i^{l-j+l'}\sqrt{(2l+1)(2j+1)(2l'+1)}
\begin{pmatrix}
l & j & l'\cr m & s & -m'
\end{pmatrix}
\begin{pmatrix}
l & j & l'\cr 0 & 0 & 0
\end{pmatrix}\, ,
\en
where
\eq
{\bf d}=\frac{2\pi}{L}\,{\bf P}\, ,\quad\quad
\eta=\frac{|{\bf p}^*|L}{2\pi}\, ,
\en
and 
\eq
Z^{{\bf d}}_{lm}(1;s)=\sum_{{\bf r}\in P_d}
\frac{{\cal Y}_{lm}({\bf r})}{{\bf r}^2-\eta^2}\, ,\quad
P_d=\{{\bf r}=\mathbb{R}^3\,|\,r_\parallel=\gamma^{-1}(n_\parallel-\mu_1|{\bf d}|),~
{\bf r}_\perp={\bf n}_\perp,\quad {\bf n}\in\mathbb{Z}^3\}\, .
\en
Note that $Z^{{\bf d}}_{lm}(1;s)$ is a function of $s$ and not merely
$\eta^2$, as in the rest frame. This happens because the kinematical factor 
$\gamma$ depends on $s$.

The finite-volume spectrum is determined by the pole positions of the scattering
matrix. The poles emerge when the determinant of the system of linear
equations~(\ref{eq:LS-L}) vanishes. Taking into account 
Eqs.~(\ref{eq:unitarity},\ref{eq:M}), the equation determining the
the energy spectrum can be written in the following form:
\eq\label{eq:luesher-moving}
\det\biggl(\delta_{ll'}\delta_{mm'}-\tan\delta_l(s){\cal M}_{lm,l'm'}(s;{\bf P})\biggr)=0\, .
\en
\begin{sloppypar}
This is L\"uscher's equation in a moving frame, or the Gottlieb-Rummukainen formula 
(see Refs.~\cite{gottlieb,Sachrajda,Davoudi,Fu}). 
It can be also shown that, in the large-$L$ limit, the equations obtained
in Ref.~\cite{Oset-movingframe} reduces to Eq.~(\ref{eq:luesher-moving}),
if all partial waves, except the S-wave, are neglected. Using discrete
symmetries, the system of linear equations~(\ref{eq:luesher-moving}),
 that couples all partial waves, can be partially diagonalized. We do 
not, however, address this problem here. A full-fledged  group-theoretical
analysis of the Gottlieb-Rummukainen formula with the inclusion of the
spin of the particles forms the subject of a separate investigation~\cite{Bob}.
\end{sloppypar}

The equation~(\ref{eq:luesher-moving}) enables one to extract the 
scattering phase shift from the measured energy spectrum on the lattice.
In order to extract a resonance pole position in the complex plane
from the phase, additional effort is needed. For example, one could assume
that the {\em effective range expansion is valid up to the resonance
energy.} This assumption works well, e.g., for the physical
$\Delta$-resonance.
The effective-range expansion for the scattering phase shift is written as:
\eq
p^{2l+1}\cot\delta_l(s)=-\frac{1}{a_l}+\frac{1}{2}\,r_lp^2+O(p^4)\, ,\quad\quad
p^2=\frac{\lambda(s,m_1^2,m_2^2)}{4s}\, .
\en
This means that the lattice data allow one to determine the scattering length
$a_l$, the effective range $r_l$, etc. The pole position $p_R$
(on the second sheet)
is then determined by solving an algebraic equation with known coefficients:
\eq
p_R^{2l+1}\cot\delta_l(s_R)=
-\frac{1}{a_l}+\frac{1}{2}\,r_lp_R^2+\cdots
=-ip_R^{2l+1}\, .
\en
It should be stressed that, in order to justify the application
of this procedure, the data should cover the energy range where the resonance
mass is located. There exist alternative strategies, which may be applied,
if the use of the effective-range expansion is questionable. However,
the present paper is mainly focused on the study of the resonance matrix
elements. In order to make the conceptual discussion of this issue as
transparent as possible, below we restrict ourselves
to the situation where the effective-range expansion can be used without problems.

\section{L\"uscher-Lellouch formula for the scalar form factor\\ from the non-relativistic EFT}
\label{sec:LL}

Before investigating the resonance matrix elements, we consider the simpler
problem for matrix elements of  stable states and re-derive the
L\"uscher-Lellouch formula~\cite{Lellouch} in an arbitrary moving frame
within the non-relativistic EFT. To ease notations, we treat the  equal mass
case $m_1=m_2=m$ here, albeit the formalism can be straightforwardly
generalized to the unequal-mass case\footnote{We have in mind, e.g., 
the calculation of the pion form factor.}.
As an example, we consider the  (scalar) form factor in the time-like
region. 
In order to study the form factor, the non-relativistic Lagrangian in
Eq.~(\ref{eq:lag}) should be equipped by the part that describes the 
interaction with the external field $A(x)$. This part 
of the Lagrangians takes the form
\eq\label{eq:lag-em}
{\cal L}_{\sf  A}=eA(x) j(x)=eA\biggl\{\Phi_1^\dagger\Phi_2^\dagger
+D_1\biggr[\Phi_1^\dagger(\Phi_2^\dagger)_\mu^\mu
+2(\Phi_1^\dagger)^\mu(\Phi_2^\dagger)_\mu
+(\Phi_1^\dagger)^\mu_\mu\Phi_2^\dagger\biggr]
+\cdots\biggr\}+\mbox{h.c.}
\, ,
\en
where (cf with Eq.~(\ref{eq:calP}))
\eq
(\Phi_i^\dagger)^{\mu\cdots}_{\nu\cdots}=({\cal P}_i^\dagger)^{\mu}\cdots
({\cal P}_i^\dagger)_{\nu}\cdots\Phi_i^\dagger\, ,
\en
and the low-energy constants $e,D_1,\cdots$ describe the coupling of the
field $A(x)$ to $\Phi_{1,2}$
(note that a similar approach 
to the electroweak matrix elements in the two-nucleon
sector of QCD was adopted in Ref.~\cite{Savage-formfactor}).

Define now the operators
\eq\label{eq:On}
{\cal O}(x_0;{\bf P},{\bf k})&=&\int_{-L/2}^{L/2}d^3{\bf x}\,d^3{\bf y}\,
e^{-\frac{i}{2}\,{\bf P}({\bf x}+{\bf y})-i{\bf k}({\bf x}-{\bf y})}\,
\Phi_1(x_0,{\bf x})\Phi_2(x_0,{\bf y})\, ,
\nonumber\\[2mm]
{\cal O}^\dagger(x_0;{\bf P},{\bf k})&=&\int_{-L/2}^{L/2}d^3{\bf x}\,d^3{\bf y}\,
e^{\frac{i}{2}\,{\bf P}({\bf x}+{\bf y})+i{\bf k}({\bf x}-{\bf y})}\,
\Phi_1^\dagger(x_0,{\bf x})\Phi_2^\dagger(x_0,{\bf y})\, ,
\nonumber\\[2mm]
&&{\bf P}=\frac{2\pi}{L}\,{\bf m}\, ,\quad\quad
{\bf k}=\frac{2\pi}{L}\,\biggr({\bf n}+\frac{1}{2}\,{\bf m}\biggr)\,
,\quad\quad
{\bf m},{\bf n}\in\mathbb{Z}^3\, ,
\en
and consider the following matrix element in  Euclidean space for
$x_0>y_0$:
\eq\label{eq:sumn}
\langle 0|{\cal O}(x_0;{\bf P},{\bf k}){\cal O}^\dagger(y_0;{\bf P},{\bf k})|0\rangle
=\sum_n|\langle0|{\cal O}(0;{\bf P},{\bf k})|E_n({\bf P})\rangle|^2e^{-E_n(x_0-y_0)}\, ,
\en
where the $E_n=E_n({\bf P})$  denote the energy eigenvalues for the eigenstates with 
total momentum ${\bf P}$.

\begin{figure}[t]
\begin{center}
\includegraphics[width=12.cm]{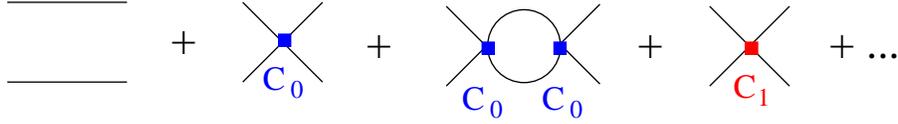}
\end{center}
\caption{Diagrams contributing to the matrix element on the l.h.s of
Eq.~(\ref{eq:sumn})
in perturbation theory. The first diagram corresponds to the free propagation
of the fields $\Phi_{1,2}$.}
\label{fig:diagrams-O}
\end{figure}
    
Note that in the non-relativistic EFT the above matrix element can be
calculated in perturbation theory. The pertinent diagrams are shown
in Fig.~\ref{fig:diagrams-O}. Using the Euclidean-space propagator
in the non-relativistic EFT
\eq
\langle 0|T\Phi_i(x)\Phi_i^\dagger(y)|0\rangle=
\int\frac{dp_0}{2\pi}\,\frac{1}{L^3}\sum_{\bf p}
\frac{e^{ip_0(x_0-y_0)+i{\bf p}({\bf x}-{\bf y})}}{2w({\bf p})(w({\bf p})+ip_0)}\, ,\quad\quad
w({\bf p})=\sqrt{m^2+{\bf p}^2}\, ,
\en  
for this matrix element we get:
\eq\label{eq:sumup}
\langle 0|{\cal O}(x_0;{\bf P},{\bf k})
{\cal O}^\dagger(y_0;{\bf P},{\bf  k})|0\rangle
=L^3\int\frac{dP_0}{2\pi}\,
e^{iP_0(x_0-y_0)}\,\hspace*{5.8cm}
\nonumber\\[2mm]
\times
\biggl\{\frac{-iL^3}{4w_1({\bf k})w_2({\bf k})(P_0-i(w_1({\bf
    k})+w_2({\bf k})))}
-\frac{T}{(4w_1({\bf k})w_2({\bf k}))^2(P_0-i(w_1({\bf
    k})+w_2({\bf k})))^2}\biggr\},
\en
where
\eq
w_1({\bf k})=
\sqrt{m^2+\biggl(\frac{\bf P}{2}+{\bf k}\biggr)^2}\, ,\quad\quad
w_2({\bf k})=
\sqrt{m^2+\biggl(\frac{\bf P}{2}-{\bf k}\biggr)^2}\, ,
\en
and $T$ is the  forward 
scattering amplitude of the particles 1 and 2 in the moving
frame (see Fig.~\ref{fig:diagrams-O}):
\eq
T&=&C_0+C_0^2\frac{1}{L^3}\sum_{\bf l}
\frac{1}{4w_1({\bf l})w_2({\bf l})
(w_1({\bf l})+w_2({\bf l})
+iP_0)}+\cdots\, .
\nonumber\\[2mm]
&=&C_0+C_0^2\frac{p^*}{8\pi^{5/2}\sqrt{s}\gamma\eta}\,
Z_{00}^{\bf  d}(1;s)+\cdots\, ,
\nonumber\\[2mm]
&&s=-(P_0^2+{\bf P}^2)\, ,\quad 
\gamma=\biggl(1+\frac{{\bf P}^2}{P_0^2}\biggr)^{-1/2}\, ,\quad
p^*=\sqrt{\frac{s}{4}-m^2}\, ,\quad
\eta=\frac{p^*L}{2\pi}\, ,
\en
where we have used Eqs.~(\ref{eq:M-ini},\ref{eq:M}), and where we have
retained only the S-wave contribution in the scattering matrix in
order to simplify the discussion of the scalar form factor.
Using
Eqs.~(\ref{eq:VC0},\ref{eq:unitarity}), the tree-level and bubble diagrams
in Fig.~\ref{fig:diagrams-O} can be summed up to all orders. The result
on the energy shell is given by
\eq\label{eq:Tsumup}
T=\frac{8\pi\sqrt{s}}{p^*\cot\delta(s)+p^*\cot\phi^{\bf d}(s)}\,
,\quad\quad
\tan\phi^{\bf d}(s)=-\frac{\pi^{3/2}\eta\gamma}{Z_{00}^{\bf d}(1;s)}\, ,
\en
where $\delta(s)=\delta_0(s)$ denotes the S-wave phase shift.

\begin{figure}[t]
\begin{center}
\includegraphics[width=10.cm]{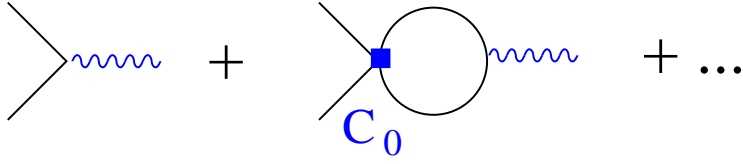}
\end{center}
\caption{Diagrams contributing to the vertex function.
The wiggly line corresponds to the external field $A(x)$.}
\label{fig:diagrams-vertex}
\end{figure}

The eigenvalues are determined from the Gottlieb-Rummukainen 
equation (see section~\ref{sec:moving}):
\eq
\delta(s)=-\phi^{\bf d}(s)+\pi n
\, ,\quad\quad s=s_n\, ,\quad {\bf P}~\mbox{fixed.}
\en
The quantity $T$ defined by Eq.~(\ref{eq:Tsumup}) has poles at
real values of $s=s_n$, i.e., at $P_0=P_{0n}=iE_n$ where $E_n=E_n({\bf P})
=\sqrt{s_n+{\bf P}^2}$. In the vicinity
of this pole, the quantity $T$ behaves as:
\eq\label{eq:vicinity}
T\to\frac{32\pi\sin^2\delta(s_n)}{\delta'(s)+(\phi^{\bf d}(s_n))'}\,
\frac{\sqrt{s_n}}{E_n}\,
\frac{1}{E_n+iP_0}+\mbox{regular terms}\, ,
\en
where the derivative is taken with respect to the variable $p^*$.
Substituting now this expression into Eq.~(\ref{eq:sumup}), performing
the integral over $P_0$ and taking into account the fact that
the ``free'' poles at $P_0=i(w_1({\bf k})+w_2({\bf k}))$ cancel in the integrand,
the final expression for the matrix element in Eq.~(\ref{eq:sumup}) 
for $x_0-y_0>0$ reads: 
\eq
\langle 0|{\cal O}(x_0;{\bf P},{\bf k})
{\cal O}^\dagger(y_0;{\bf P},{\bf  k})|0\rangle
&=&L^3\sum_n\frac{32\pi\sin^2\delta(s_n)}{\delta'(s)+(\phi^{\bf d}(s_n))'}\,
\frac{\sqrt{s_n}}{E_n}\,
\nonumber\\[2mm]
&\times&\frac{e^{-E_n(x_0-y_0)}}{(4w_1({\bf k})w_2({\bf k}))^2
(E_n-w_1({\bf k})-w_2({\bf k}))^2}\, .
\en
Comparing this expression with Eq.~(\ref{eq:sumn}), one reads off:
\eq\label{eq:abs-O}
 |\langle0|{\cal O}(0;{\bf P},{\bf k})|E_n({\bf P})\rangle|
&=&
L^{3/2}\biggl(\frac{32\pi\sin^2\delta(s_n)}{|\delta'(s)+(\phi^{\bf d}(s_n))'|}\,
\frac{\sqrt{s_n}}{E_n}\biggr)^{1/2}\,
\nonumber\\[2mm]
&\times& \frac{1}{4w_1({\bf k})w_2({\bf k})}\,
\frac{1}{|E_n-w_1({\bf k})-w_2({\bf k})|}\, .
\en
Next, we turn to the determination of the form factor in the time-like region.
To this end, we have to consider the amplitude of  pair
creation from the vacuum in the presence of an external field
$A(x)$, at the first order in the coupling $e$. This matrix element is described by
\eq\label{eq:form-sec}
\langle 0|{\cal O}(x_0,{\bf P},{\bf k})\,{\cal L}_{\sf A}(0)|0\rangle
=eA(0) F[{\bf k},{\bf P};x_0]\, ,\quad\quad
x_0>0\, .
\en
We evaluate the quantity $F$ in perturbation theory.
The pertinent diagrams are shown in Fig.~\ref{fig:diagrams-vertex}. Summing
up all bubbles yields: 
\eq\label{eq:Fmu}
 F[{\bf k},{\bf P};x_0]=\bar F(t)
\int\frac{dP_0}{2\pi i}\frac{e^{iP_0x_0}}{4w_1({\bf k})w_2({\bf k})
(P_0-i(w_1({\bf k})+w_2({\bf k})))}\,
\frac{p^*\cot\delta(s)}{p^*\cot\delta(s)+p^*\cot\phi^{\bf d}(s)}\, ,
\en
where the quantity $\bar F(t)$  can be read off the
Lagrangian in Eq.~(\ref{eq:lag-em}) at tree level       
\eq
\bar F(t)&=&1+D_1t+O(t^2)\, ,\quad\quad 
t=(k_1+k_2)^2\, ,
\nonumber\\[2mm]
k_1^\mu&=&\biggl(w_1({\bf k}),\frac{\bf P}{2}+{\bf k}\biggr)\, ,\quad
k_2^\mu=\biggl(w_2({\bf k}),\frac{\bf P}{2}-{\bf k}\biggr)\, .
\en

\noindent
Using Eq.~(\ref{eq:vicinity}), we may now perform the integration over the
variable $P_0$ in Eq.~(\ref{eq:Fmu}), with the result
\eq\label{eq:Fmu1}
 F[{\bf k},{\bf P};x_0]=\bar F(t)\sum_n\frac{e^{-E_nx_0}}
{4w_1({\bf k})w_2({\bf k})(w_1({\bf k})+w_2({\bf k})-E_n)}\,
\frac{4p^*\cot\delta(s_n)\sin^2\delta(s_n)}{(\delta'(s)+(\phi^{\bf d}(s_n))')
E_n}\, .
\en  
On the other hand, the matrix element in Eq.~(\ref{eq:form-sec}) has the following
representation:
\eq\label{eq:form1}
\langle 0|{\cal O}(x_0,{\bf P},{\bf k})\,{\cal L}_{\sf A}(0)|0\rangle
=eA(0)\sum_ne^{-E_nx_0}
\langle 0|{\cal O}(0,{\bf P},{\bf k})|E_n({\bf P})\rangle
\langle E_n({\bf P})|j(0)|0\rangle\, .
\en
Using Eqs.~(\ref{eq:abs-O}), (\ref{eq:Fmu1}) and (\ref{eq:form1}), we get
\eq\label{eq:form-finite}
|\langle E_n({\bf P})|j(0)|0\rangle|=
L^{-3/2}|\bar F(t)|
\frac{p^*|\cos\delta(s_n)|}{(2\pi\sqrt{s}E_n)^{1/2}}\,
\frac{1}{|\delta'(s)+(\phi^{\bf d}(s_n))'|^{1/2}}\, .
\en
This is the expression of the matrix element in a finite volume. It should
be compared with its counterpart in the infinite volume, which is obtained 
by using Watson's theorem: 
\eq\label{eq:form-infinite}
\langle k_1,k_2;\mbox{\sf out}|j(0)|0\rangle=F(t)\, ,\quad\quad
|F(t)|= |\bar F(t)\cos\delta(s)|\, .
\en
From Eqs.~(\ref{eq:form-finite}) and (\ref{eq:form-infinite}) we finally get:
\eq\label{eq:LL-final}
|F(t)|^2
=|L^{3/2}\langle E_n({\bf P})|j(0)|0\rangle|^2
\frac{2\pi\sqrt{s}E_n}{(p^*)^2}\,
|\delta'(s)+(\phi^{\bf d}(s_n))'|\, .
\en
This expression allows one to extract the absolute value of a scalar
form factor in the time-like region from the measured matrix element
$\langle E_n({\bf P})|j(0)|0\rangle$ in a finite volume. Since the phase of
this form factor, which is determined by Watson's theorem, is also measurable
on the lattice, we finally conclude that the real and imaginary parts of the
form factor can be measured on the lattice in the elastic region.

In the rest frame, the expression in Eq.~(\ref{eq:LL-final}) is similar
to the expression obtained in Ref.~\cite{Meyer}, apart from a difference in a
kinematical factor which is related to the fact that there 
a vector form factor instead of a scalar one was
considered. It can be also shown that, by using our method, one exactly
reproduces the L\"uscher-Lellouch formula in moving 
frames~\cite{deDivitiis:2004rf,Sachrajda,Christ:2005gi}.

\section{Extraction of resonance matrix elements in a finite volume}
\label{sec:vertex}

Having considered the case of the form factor in the time-like region in great
detail, we turn to the extraction of the resonance form factor. The part of
the Lagrangian that describes the interaction with the external scalar field
$A(x)$, looks now as follows:
\eq\label{eq:lbar}
\bar {\cal L}_{\sf A}=A(x)\bar j(x)=e_1A(\Phi_1^\dagger\Phi_1+\cdots)+  
e_2A(\Phi_2^\dagger\Phi_2+\cdots)+E_0A(\Phi_1^\dagger\Phi_2^\dagger\Phi_1\Phi_2
+\cdots)\, ,
\en
where $e_{1,2},E_0,\cdots$ denote low-energy couplings, and the ellipses stand
for the terms with higher derivatives. It is seen that, in general, 
 the current $\bar j(x)$
consists of  one-body currents and a two-body current, whose coupling
at lowest order is given by $E_0$. 

We make the following choice for the resonance field operators:
\eq\label{eq:On-vertex}
{\cal O}(x_0;{\bf P})&=&\int_{-L/2}^{L/2}d^3{\bf x}\,
e^{-i{\bf P}{\bf x}}\,
\Phi_1(x_0,{\bf x})\Phi_2(x_0,{\bf x})\, ,
\nonumber\\[2mm]
{\cal O}^\dagger(y_0;{\bf Q})&=&\int_{-L/2}^{L/2}d^3{\bf y}\,
e^{i{\bf Q}{\bf y}}\,
\Phi_1^\dagger(y_0,{\bf y})\Phi_2^\dagger(y_0,{\bf y})\, .
\en
The first-order scattering amplitude of the particles 1 and 2 in the external field $A(x)$
can again be calculated using perturbation theory. The pertinent diagrams
are depicted in Fig.~\ref{fig:diagrams-res}. The result of the calculation is
(cf. with Eqs.~(\ref{eq:form-sec},\ref{eq:Fmu})):

\begin{figure}[t]
\begin{center}
\includegraphics[width=14.cm]{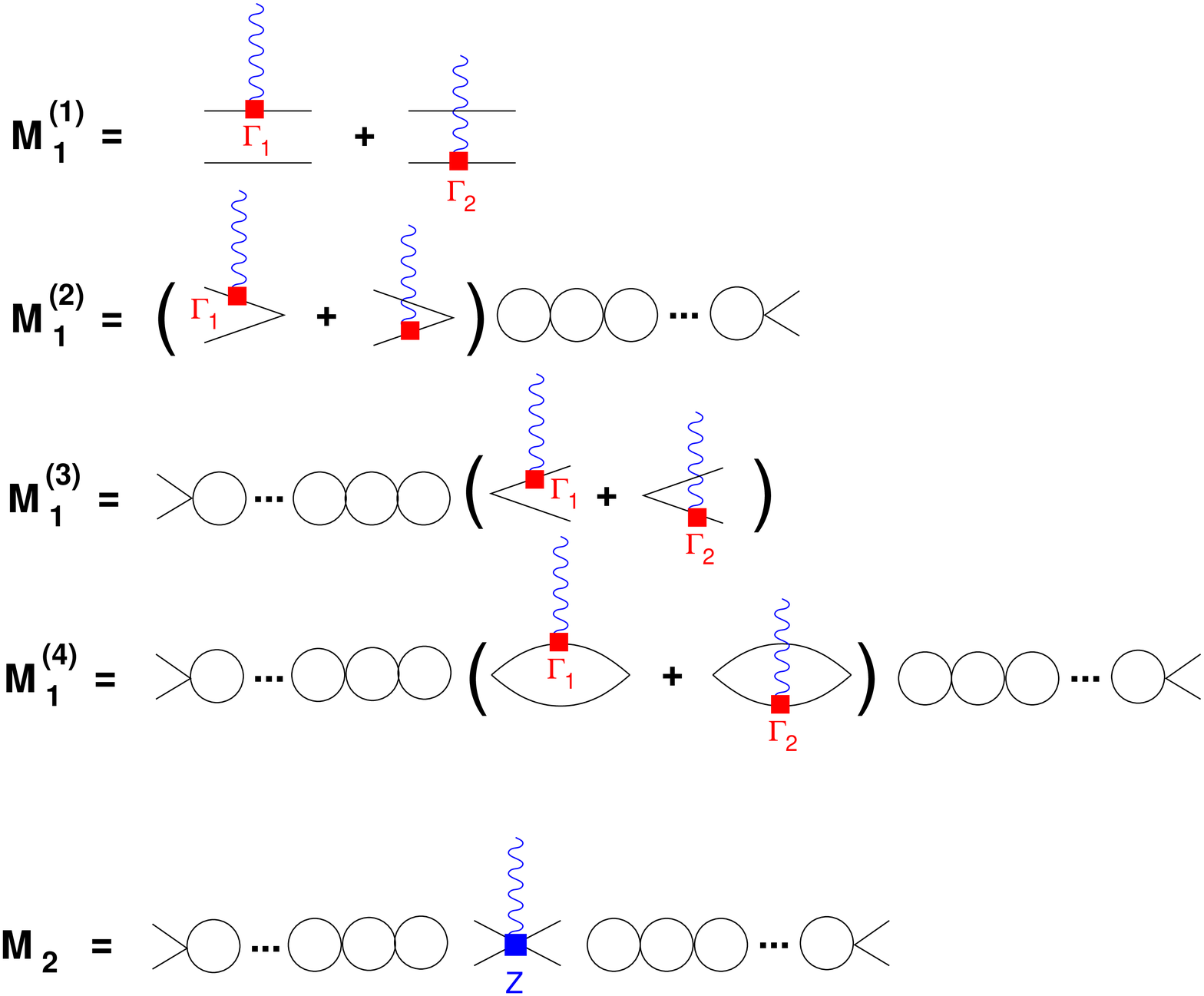}
\end{center}
\caption{The diagrams describing the quantity $M=M_1+M_2$ in
  Eq.~(\ref{eq:M-res}). $M_1$ and $M_2$ correspond to the contributions of the
  one- and two-body currents, respectively.
The wiggly line corresponds to the external field $A(x)$. 
All initial-
and final-state interactions are summed up in the bubble chains.}
\label{fig:diagrams-res}
\end{figure}
    
\eq
\langle 0|{\cal O}(x_0,{\bf P})\bar {\cal L}_{\sf A}(0)
{\cal O}^\dagger(y_0,{\bf Q})|0\rangle=A(0)\bar F({\bf P},{\bf Q},x_0,y_0)\,
,
\quad\quad
x_0>0,~y_0<0\, ,
\en
where
\eq\label{eq:M-res}
F({\bf P},{\bf Q},x_0,y_0)=
\frac{1}{L^6}\sum_{{\bf p},{\bf q}}
\int\frac{dP_0}{2\pi i}\,\frac{dQ_0}{2\pi i}\,\frac{e^{iP_0x_0}\,
M(P_0,{\bf P},{\bf p};Q_0,{\bf Q},{\bf q})\,e^{-iQ_0y_0}}
{4w_1w_2
(P_0-i(w_1+w_2))4w_1'w_2'(Q_0-i(w_1'+w_2'))}\, ,
\en
where
\eq
&&w_1=\sqrt{m^2+\biggl(\frac{\bf P}{2}+{\bf p}\biggr)}\, ,\quad
w_2=\sqrt{m^2+\biggl(\frac{\bf P}{2}-{\bf p}\biggr)}\, ,\quad
\nonumber\\[2mm]
&&w_1'=\sqrt{m^2+\biggl(\frac{\bf Q}{2}+{\bf q}\biggr)}\, ,\quad
w_2'=\sqrt{m^2+\biggl(\frac{\bf Q}{2}-{\bf q}\biggr)}\, .
\en
The diagrammatic expansion of the quantity $M$ is shown in
Fig.~\ref{fig:diagrams-res}. It consists of the contributions corresponding to
the one-body and two-body currents (see Eq.~(\ref{eq:lbar})). Retaining only
the S-wave contribution in the initial- and final-state rescattering amplitudes,
we get:
\eq
M=M_1+M_2\, ,\quad\quad
M_1=M_1^{(1)}+M_1^{(2)}+M_1^{(3)}+M_1^{(4)}\, ,
\en
where 
\eq
M_1^{(1)}&=&(2w_2)\Gamma_1\biggl(\frac{\bf P}{2}+{\bf p},
\frac{\bf Q}{2}+{\bf q}\biggr)
L^3\delta_{\frac{\bf P}{2}-{\bf p},\frac{\bf Q}{2}-{\bf q}}
+(2w_1)\Gamma_2\biggl(\frac{\bf P}{2}-{\bf p},
\frac{\bf Q}{2}-{\bf q}\biggr)
L^3\delta_{\frac{\bf P}{2}+{\bf p},\frac{\bf Q}{2}+{\bf q}}\, ,
\nonumber\\[2mm]
M_1^{(2)}&=&\frac{\Gamma_1\biggl(\dfrac{\bf P}{2}+{\bf p},
{\bf Q}-\dfrac{\bf P}{2}+{\bf p}\biggr)}
{2\tilde w_1(\tilde w_1+w_2+iQ_0)}\,S(q^*)
+\frac{\Gamma_2\biggl(\dfrac{\bf P}{2}-{\bf p},
{\bf Q}-\dfrac{\bf P}{2}-{\bf p}\biggr)}
{2\tilde w_2(\tilde w_2+w_1+iQ_0)}\,S(q^*)+\bar M_1^{(2)}\, ,
\nonumber\\[2mm]
M_1^{(3)}&=&\frac{\Gamma_1\biggl({\bf P}-\dfrac{\bf Q}{2}+{\bf q},
\dfrac{\bf Q}{2}+{\bf q}\biggr)}
{2\tilde w_1'(\tilde w_1'+w_2'+iP_0)}\,S(p^*)
+\frac{\Gamma_2\biggl({\bf P}-\dfrac{\bf Q}{2}-{\bf q},
\dfrac{\bf Q}{2}-{\bf q}\biggr)}
{2\tilde w_2'(\tilde w_2'+w_1'+iP_0)}\,S(p^*)+\bar M_1^{(3)}\, ,
\nonumber\\[2mm]
M_1^{(4)}&=&
\frac{1}{L^3}\sum_{\bf l}
\frac{S(p^*)(\Gamma_1({\bf P}-{\bf l},{\bf Q}-{\bf l})+
\Gamma_2({\bf P}-{\bf l},{\bf Q}-{\bf l}))S(q^*)}
{8w({\bf P}-{\bf l})w({\bf Q}-{\bf l})w({\bf l})
(w({\bf Q}-{\bf l})+w({\bf l})+iQ_0)(w({\bf P}-{\bf l})+w({\bf l})+iP_0)}
\nonumber\\[2mm]
&+&\bar M_1^{(4)}\, .
\en
In the above expressions, $\bar M_1^{(2)}$ and $\bar M_1^{(3)}$ do not contain
denominators linear in $Q_0$ and $P_0$, respectively, but still include the
factors $S(q^*),S(p^*)$. The quantity $\bar M_1^{(4)}$ contains at most one
energy denominator and both $S(q^*)$ and $S(p^*)$. 
These quantities emerge, because initial- and final-state rescattering
occurs, in general, off the energy shell. Further,
\eq
\tilde w_1&=&\sqrt{m^2+\biggl({\bf Q}-\frac{\bf P}{2}+{\bf p}\biggr)^2}\, ,\quad
\tilde w_2=\sqrt{m^2+\biggl({\bf Q}-\frac{\bf P}{2}-{\bf p}\biggr)^2}\,
,\quad
\nonumber\\[2mm]
\tilde w_1'&=&\sqrt{m^2+\biggl({\bf P}-\frac{\bf Q}{2}+{\bf q}\biggr)^2}\, ,\quad
\tilde w_2'=\sqrt{m^2+\biggl({\bf P}-\frac{\bf Q}{2}-{\bf q}\biggr)^2}\, ,
\en
\eq
S(p^*)&=&\frac{8\pi\sqrt{s}}{p^*\cot\delta(s)+p^*\cot\phi^{\bf d}(s)}\,,\quad 
S(q^*)=\frac{8\pi\sqrt{s}}{q^*\cot\delta(s')+q^*\cot\phi^{{\bf d}'}(s')}\, ,
\nonumber\\[2mm]
s&=&-(P_0^2+{\bf P}^2)\, ,\quad  s'=-(Q_0^2+{\bf Q}^2)\, ,
\nonumber\\[2mm]
p^*&=&\sqrt{\frac{s}{4}-m^2}\,,\quad q^*=\sqrt{\frac{s'}{4}-m^2}\,,\quad
{\bf d}=\frac{2\pi{\bf P}}{L}\,,\quad {\bf d}'=\frac{2\pi{\bf Q}}{L}\, ,
\en
 and the $\Gamma_{1,2}$ denote the tree-level interaction vertices of the external
 field $A(x)$ with the fields $\Phi_{1,2}$.

After projection onto S-waves,
the two-body current leads to the following contribution
(see Fig.~\ref{fig:diagrams-res}):
\eq
M_2&=&\frac{p^*\cot\delta(s)}{p^*\cot\delta(s)+p^*\cot\phi^{\bf d}(s)}\,
Z(iP_0,{\bf P};iQ_0,{\bf Q})\, 
\frac{q^*\cot(s')}{q^*\cot\delta(s')+q^*\cot\phi^{{\bf d}'}(s')}
\nonumber\\[2mm]
&+&\mbox{regular functions in $P_0$ or $Q_0$}, ,
\en
where the quantity $Z$ is a low-energy polynomial.

It can be straightforwardly checked that the sum of all terms 
in the integrand in Eq.~(\ref{eq:M-res})
do not have singularities at the free two-particle levels. The only
singularities are  simple poles that correspond to the energy levels in the
full theory and emerge after the summation of the bubble chains. Taking this
fact into account and performing the contour integration in the variables
$P_0,Q_0$ by using Cauchy's theorem, we get:
\eq\label{eq:M-res1}
F({\bf P},{\bf Q},x_0,y_0)&=&\frac{1}{L^6}\sum_{{\bf p},{\bf q}}\sum_{n,m}
\frac{32\pi\sin^2\delta(s_n)\sqrt{s_n}e^{-E_nx_0}}
{4w_1w_2E_n(w_1+w_2-E_n)(\delta'(s_n)+(\phi^{\bf d}(s_n))')}\,
V_{nm}({\bf P};{\bf Q})
\nonumber\\[2mm]
&\times&
\frac{32\pi\sin^2\delta(s_m)\sqrt{s_m}e^{E_my_0}}
{4w_1'w_2'E_m(w_1'+w_2'-E_m)(\delta'(s_m)+(\phi^{\bf d}(s_m))')}\, ,
\en
where
\eq\label{eq:Vnm-def}
V_{nm}({\bf P};{\bf Q})&\!=\!&
\frac{1}{L^3}\sum_{\bf l}
\frac{\Gamma_1({\bf P}-{\bf l},{\bf Q}-{\bf l})+
\Gamma_2({\bf P}-{\bf l},{\bf Q}-{\bf l})}
{8w({\bf P}-{\bf l})w({\bf Q}-{\bf l})w({\bf l})
(w({\bf P}-{\bf l})+w({\bf l})-E_n)(w({\bf Q}-{\bf l})+w({\bf l})-E_m)}
\nonumber\\[2mm]
&\!+\!&
\frac{p^*_n\cot\delta(s_n)}{8\pi\sqrt{s_n}}\,
Z(E_n,{\bf P};E_m,{\bf Q})
\frac{q^*_m\cot\delta(s_m)}{8\pi\sqrt{s_m}}
\en
with $p^*_n=p^*(s=s_n)$, $q^*_m=q^*(s'=s_m)$.
In Eq.~(\ref{eq:M-res1}), the Gottlieb-Rummukainen equation will be further 
used to remove the summations over ${\bf p},{\bf q}$:
\eq
\frac{1}{L^3}\sum_{\bf p}\frac{1}{4w_1w_2(w_1+w_2-E_n)}
=\frac{p_n^*\cot\delta(s_n)}{8\pi\sqrt{s_n}}\, .
\en
On the other hand, inserting a full set of the eigenstates of the full
Hamiltonian, we get
\eq
F({\bf P},{\bf Q},x_0,y_0)&=&\sum_{n,m}\langle 0|{\cal O}(0;{\bf P})|
E_n({\bf P})\rangle e^{-E_nx_0}\langle E_n({\bf P})|\bar j(0)|E_m({\bf Q})\rangle
\nonumber\\[2mm]
&\times&e^{E_my_0}\langle E_m({\bf Q})|{\cal O}^\dagger(0;{\bf Q})|0\rangle\, .
\en
Further, by using perturbation theory, it is straightforward to show that
\eq
|\langle 0|{\cal O}(0;{\bf P})|E_n({\bf P})\rangle|^2=
L^3\,\frac{\cos^2\delta(s_n)}{\delta'(s_n)+(\phi^{\bf d}(s_n))'}\,
\frac{(p_n^*)^2}{2\pi E_n({\bf P})\sqrt{s_n}}\, .
\en
Taking $n=m$, we readily obtain:
\eq\label{eq:matrix-finite}
\langle E_n({\bf P})|\bar j(0)|E_n({\bf Q})\rangle
=\frac{(4\sin\delta(s_n))^2}{\delta'(s_n)+(\phi^{\bf d}(s_n))'}\,
\frac{2\pi\sqrt{s_n}}{L^3\sqrt{E_n({\bf P})E_n({\bf Q})}}\,
V_{nn}({\bf P};{\bf Q})\, .
\en
Independently, one may extract the resonance matrix element in the
infinite-volume non-relativistic EFT by using the procedure described in the
introduction. The result is given by:
\eq\label{eq:matrix-infinite}
\langle {\bf P}|\bar j(0)|{\bf Q}\rangle
=B_R\,V^\infty({\bf P};{\bf Q})\, ,
\quad\quad s,s'\to s_R=4(m^2+p_R^2)\, ,
\en
where
\eq
\frac{8\pi\sqrt{s}}{p\cot\delta(p)-ip}\to\frac{B_R}{s_R-s}\, ,\quad\quad
B_R=-\frac{64\pi\sqrt{s_R}p_R}{2p_Rh'(p_R^2)-i}\, ,
\en
and
\eq
h(p^2)=p\cot\delta(p)=-\frac{1}{a}+\frac{1}{2}\,rp^2+\cdots\, ,
\en
and $V^\infty$ is obtained from $V_{nm}$ through $E_n\to P_0$, 
$E_m\to Q_0$, $s,s'\to s_R$ and further replacing the discrete sum 
by integration over the variable ${\bf l}$.
 
At this stage, we can visualize the problem inherent to the extraction of the
resonance matrix elements. On the lattice, one may measure 
the quantity $\langle E_n({\bf P})|\bar j(0)|E_n({\bf Q})\rangle$ and extract
the quantity $V_{nn}$ through Eq.~(\ref{eq:matrix-finite}). If we were
dealing with a stable bound state, in the infinite volume 
$V_{nn}\to V^\infty$, up to  exponentially small corrections. Multiplying
with the pertinent bound-state renormalization factor, we would directly
arrive at the matrix element of the current $\bar j(0)$, sandwiched between 
the stable  
bound-state vectors. However, we are dealing with a resonance and not with a
stable bound state. This means that:
\begin{itemize}
\item[i)] No single $E_n$ corresponds to a resonance. We have to formulate a
  procedure for the analytic continuation of the matrix elements into the
  complex plane.

\item[ii)]
The quantity $V_{nn}$ does not have a well-defined limit as $L\to\infty$ and
$E_n$ above the two-particle threshold. The 1-loop diagram with an external
field, which contributes to the $M_1^{(4)}$, is the culprit. On the contrary, the
contribution from the two-body current, $Z$, is a low-energy polynomial and
does not cause any problem.
\end{itemize}

In the following sections, we shall explicitly address both of these problems.

\section{Analytic continuation and fixed points}
\label{sec:fixed}

In order to avoid kinematical complications, let us first consider the
form factor at a zero momentum transfer ${\bf P}={\bf Q}=0$. The quantity
$V_{nn}$ is then a function of a single variable $p=\sqrt{E_n^2/4-m^2}$. The
  questions can be now formulated as follows:
\begin{itemize}
\item[i)] 
How does one perform the analytic continuation $p\to p_R$ in the quantity
$V_{nn}(p)$?
\item[ii)]
How does one perform the infinite volume limit $L\to\infty$?

\end{itemize}

We shall see below that these two questions are intimately related. 

Let us imagine for a moment that the contribution from the loop diagrams
vanishes, so that the quantity $V_{nn}$ is given by the two-body current
diagram $Z$ only. Then, the answers to the above equations are trivial. The
quantity $Z$ is a polynomial in the variable $p^2$: $Z=Z_0+Z_1p^2+\cdots$.
So, one has to first fit the coefficients $Z_0,Z_1,\cdots$ to the
lattice data, and then simply substitute $p^2\to p_R^2$. The result gives
the analytic continuation $Z(p^2)\to Z(p_R^2)$.
Moreover, since $Z(p^2)$ is
$L$-independent, so is $Z(p_R^2)$, and the final result does not depend on the
energy level $n$ we started from.

Let us now see what changes when the one-body current contribution is also
included. To this end, we first study the analytic continuation of the L\"uscher
equation into the complex plane. 
To ease notation, we restrict ourselves 
to S-waves and write down the equation (in the CM frame) 
in the following form:
\eq\label{eq:luescher-complex}
\frac{h(p^2)}{p}=
\cot\delta(p)=\frac{1}{\pi^{3/2} \eta}\,Z_{00}(1,\eta^2)\, .
\en
On the real axis,
\eq\label{eq:eta-L}
\eta=\frac{pL}{2\pi}\, ,
\en
and Eq.~(\ref{eq:luescher-complex})
 determines the energy levels given the scattering phase (or vice versa).
Let us now look for solutions of this
equation for  {\em complex} values of $p$. The quantity $h(p^2)$ is a
low-energy polynomial in $p^2$, so the analytic continuation is
trivial. Furthermore, the function $Z_{00}(1,\eta^2)$ is a meromorphic
function of the variable $\eta^2$. Thus, for
any given complex value of $p$, the solutions of Eq.~(\ref{eq:luescher-complex})
determine the trajectories $\eta_n(p),~n=0,1,\ldots$\,, 
in the complex plane (we remind the
reader that the solutions are not unique). As $p\to p_R$ in the $p$-plane,
    $\eta_n(p)\to\eta_{nR}$ in the $\eta$-plane and Eq.~(\ref{eq:eta-L})
becomes a relation that defines $L$. Our first task is to find 
all $\eta_{nR}$.

It is instructive to begin from the 1+1-dimensional case~\cite{1+1}. The
counterpart of Eq.~(\ref{eq:luescher-complex}) in this case reads:
\eq
\cot\delta(p)=-\cot\pi\eta\, .
\en
The solution of this equation with respect to $\eta$ reads:
\eq
\eta=-\frac{i}{2\pi}\,\ln\frac{-1+ix}{1+ix}\, ,\quad\quad x=\cot\delta(p)\, .
\en
On the resonance position, we have $p\to p_R$ and $\cot\delta(p)\to -i$. Writing
$x=-i+\epsilon$, we get
\eq
\eta\sim \frac{1}{2\pi}\arg i\epsilon-\frac{i}{2\pi}\,\ln\frac{|\epsilon|}{2}
+O(\epsilon)\, ,\quad\mbox{as}~p\to p_R\, .
\en
If we exclude those paths
connecting $p$ and $p_R$ in the $p$-plane, which 
wind around $p_R$ infinitely many times, then
\eq\label{eq:fixed-infinite}
\lim_{p\to p_R}\mbox{Re}\,\eta(p)<\infty\, ,\quad\quad
\mbox{Im}\,\eta(p)\to -\infty,\quad\mbox{as}~p\to p_R\, .
\en
Recalling the definition of the variable $\eta$ (see
Eq.~(\ref{eq:eta-L})), one may interpret the above result (in a
loose sense) as the equivalence of the mass-shell limit for a
resonance ($p\to p_R$) and the
infinite-volume limit. The same is true for a stable bound state: its energy
is volume-independent up to  exponentially small corrections, so the walls
can be safely moved to infinity. Our result shows that the same statement
holds for a resonance pole (in the 1+1 dimensional case). On the contrary, the
discrete spectrum above the two-particle threshold is determined by the
presence of the walls. If one moves the walls to infinity ($L\to\infty$), each
given energy level collapses toward threshold. The spectrum becomes continuous
in this limit. 

What does change in the 3+1 dimensional case? There are so-called {\em finite
  fixed points} with $|\eta_{nR}|<\infty$, in addition to the {\em fixed
  points at infinity} which are given by Eq.~(\ref{eq:fixed-infinite}). In
order to see this, we provide below a numerical solution of
Eq.~(\ref{eq:luescher-complex}) (an analytical solution is not available in
the $3+1$-dimensional case).  

\begin{figure}[t]
\begin{center}
\includegraphics[width=10.cm]{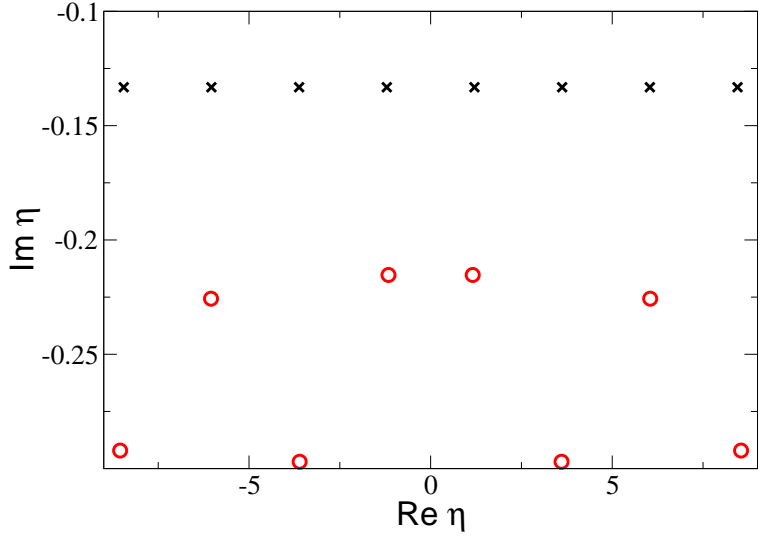}
\end{center}
\caption{Positions of the finite fixed points in the complex $\eta$-plane.
The crosses and circles denote $\eta_{nR}^{(0)}$ and $\eta_{nR}$,
respectively. The plot is symmetric with respect to the imaginary axis.}
\label{fig:fixed}
\end{figure}

The fixed points are the solutions of the equation
\eq\label{eq:fixed-ini}
Z_{00}(1;\eta^2)+i\pi^{3/2}\eta=0\, .
\en
If $\mbox{Im}\,\eta<0$, one may use the following representation of the
zeta-function:
\eq\label{eq:Poisson}
Z_{00}(1;\eta^2)=\pi^{3/2}\eta\,\biggl\{-i+\sum_{|{\bf n}|\neq 0}
\frac{1}{2\pi\eta|{\bf n}|}\,e^{-2\pi i\eta |{\bf n}|}\biggr\}\, ,\quad\quad
{\bf n}\in\mathbb{Z}^3\, .
\en
By using Eq.~(\ref{eq:Poisson}), Eq.~(\ref{eq:fixed-ini}) can be rewritten
as 
\eq\label{eq:exact}
6e^{-2\pi i\eta}+\frac{12}{\sqrt{2}}\,e^{-2\pi i\sqrt{2}\eta}+\Sigma(\eta)=0\,
,
\quad\quad
\Sigma(\eta)=\sum_{|{\bf n}|\geq \sqrt{3}}
\frac{1}{|{\bf n}|}\,e^{-2\pi i\eta |{\bf n}|}\, .
\en
This equation has infinitely many solutions. In order to verify this
statement, first assume that $\Sigma(\eta)=0$. In this approximation, there
exists a tower of finite fixed poles parameterized as
\eq
\eta_{nR}^{(0)}=\frac{n+\frac{1}{2}}{\sqrt{2}-1}-\frac{i\ln
  2}{4\pi(\sqrt{2}-1)}\, ,\quad\quad n=-\infty,\cdots, -1,0,1,\cdots,\infty\, .
\en 
Finally, the equation (\ref{eq:exact}) can be rewritten as
\eq\label{eq:iterations}
\eta=\frac{n+\frac{1}{2}}{\sqrt{2}-1}-\frac{i\ln
  2}{4\pi(\sqrt{2}-1)}-\frac{i}{2\pi(\sqrt{2}-1)}\,
\ln\biggl(1+\frac{\sqrt{2}}{12}\,e^{2\pi i\sqrt{2}\eta}\,\Sigma(\eta)\biggr)\, .
\en
This equation can be easily solved by iteration, starting from 
$\eta=\eta_{nR}^{(0)}$. Note that the series for $\Sigma(\eta)$ contains
exponentially suppressed terms and converges very fast. So, truncating the sum
at some $|{\bf n}|=n_{\sf max}$ can be justified. The numerical solutions
indeed exist and are shown in Fig.~\ref{fig:fixed}.

{\it Enter the culprit.} What are the implications of the above result for the
calculation of resonance matrix elements? Consider a simplified expression for
$V_{nn}({\bf 0},{\bf 0})$ in Eq.~(\ref{eq:Vnm-def}), setting $\Gamma_1=1$,
$\Gamma_2=0$ and $Z=0$ (the low-energy polynomials in the numerator do not
alter the analytic properties of the diagram we are interested in, and the
part containing $Z$ is trivial and was considered already). All we have to
consider is the expression
\eq\label{eq:two-terms}
V_{nn}({\bf 0},{\bf 0})&=&\frac{1}{L^3}\sum_{\bf l}
\frac{1}{(2w({\bf l}))^3(2w({\bf l})-E_n)^2}=\frac{d}{dE}\biggl(
\frac{p}{8\pi E^2}\,
\frac{Z_{00}(1;\eta^2)}{\pi^{3/2}\eta}\biggr)\biggr|_{E=E_n}+\cdots
\nonumber\\[2mm]
&=&-\biggl(\frac{m^2-p^2}{8\pi E^3p^2}\,p\cot\phi(\eta)-\frac{1}{32\pi Ep}\,
(1+\cot^2\phi(\eta))\,\eta\phi'(\eta)
\biggr)\biggr|_{E=E_n}+\cdots\, ,
\nonumber\\[2mm]
&=&\biggl(\frac{m^2-p^2}{8\pi E^3p^2}\,p\cot\delta(p)+\frac{1}{32\pi Ep}\,
(1+\cot^2\delta(p))\,\eta\phi'(\eta)
\biggr)\biggr|_{E=E_n}+\cdots\, ,
\en
where $\phi(\eta)=\phi^{{\bf d}={\bf 0}}(E^2)$, $E=2\sqrt{m^2+p^2}$,
 and the ellipses stand for the
terms which vanish exponentially with $L$. In the last line of 
Eq.~(\ref{eq:two-terms}), L\"uscher's equation
$\cot\phi(\eta)=-\cot\delta(p)$ was used. 
The quantity $V_{nn}({\bf 0},{\bf 0})$ is a function of the variable
$p$, so one can write  $V_{nn}({\bf 0},{\bf 0})=V_{nn}(p)$.
It is now legitimate to ask
how the analytic continuation of the above expression in $p$ is performed and
what is the result of this continuation. 
The expression in Eq.~(\ref{eq:two-terms}) consists of two terms. 
It can be verified directly that
the first
term is a low-energy polynomial in $p^2$ (up to a trivial
 overall factor $p^{-2}$). The analytic
continuation of this term is straightforward and leads to
\eq
\frac{m^2-p^2}{8\pi E^3 p^2}\,p\cot\delta(p)\to 
-i \frac{m^2-p_R^2}{8\pi  s_R^{3/2} p_R}\, ,\quad\quad
\mbox{as}~ p\to p_R\, .
\en
It is easy to check that this result exactly coincides with the result for
the loop diagram calculated in the infinite volume
 (i.e., replacing summation
by integration in Eq.~(\ref{eq:two-terms})), on the second sheet. Consequently, if the second term,
continued to $p=p_R$, vanishes, the analytic continuation of the whole vertex
diagram to the pole on the second sheet will yield the same vertex evaluated
in the infinite volume. This would be the statement that we are after.

Let us assume for a moment that it is possible to find a procedure to perform
such an analytic continuation in the second term of
Eq.~(\ref{eq:two-terms}).
We choose some path in the complex $p$-plane
approaching the pole at $p=p_R$.  Suppose first that, moving along this path,
the variable $\eta=\eta(p)$ approaches the infinite fixed point
$|\mbox{Re}\,\eta_{nR}|<\infty$, $\mbox{Im}\,\eta_{nR}\to -\infty$.
Using the representation for the zeta-function given in
Eq.~(\ref{eq:Poisson}), it can be easily checked that the second term in
Eq.~(\ref{eq:two-terms}) indeed vanishes if $\eta$ tends to the infinite
fixed point.

\begin{figure}[t]
\begin{center}
\includegraphics[width=10.cm]{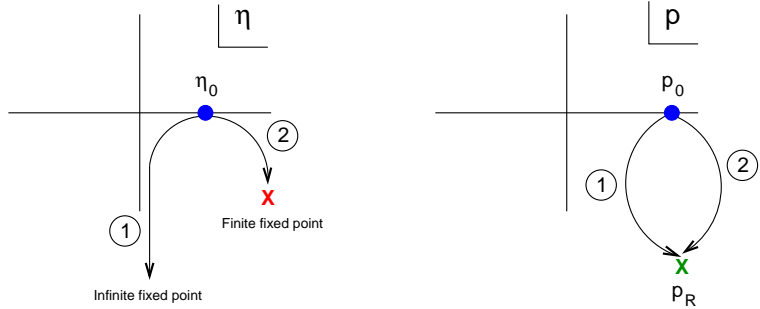}
\end{center}
\caption{Two different paths in the $\eta$-plane and corresponding paths in
  the $p$-plane. It is seen that the paths converge to the same point $p_R$ in
the $p$-plane.}
\label{fig:eta-p}
\end{figure}

\noindent
Imagine now a path
that ends at a finite fixed point. Parameterizing this path as
\eq 
\cot\phi(\eta)+i=\lambda(\eta-\eta_{nR})+O((\eta-\eta_{nR})^2)\, ,
\en
where $\lambda$ is a finite complex constant. Then, in the vicinity of 
 the fixed point,
\eq
\phi(\eta)\propto\ln(\eta-\eta_{nR})+\mbox{const}+O(\eta-\eta_{nR})\, ,
\quad\quad
\phi'(\eta)\propto\frac{1}{\eta-\eta_{nR}}+\mbox{const}+O(\eta-\eta_{nR})\, .
\en
From the above equations it is evident that the product 
$\eta(1+\cot^2\phi(\eta))\phi'(\eta)$, rather than vanishing, tends to a
constant at the finite fixed point. In other words, if during the analytic
continuation, the variable $\eta$ gets caught by a finite fixed point, the
result of the analytic continuation is different from the vertex function in
the infinite volume and one is in trouble.

In order to understand this result better, let us consider some point $\eta_0$
on the real axis and two paths, connecting $\eta_0$ to an infinite and
to a finite fixed points, respectively (see Fig.~\ref{fig:eta-p}). These paths
are mirrored by pertinent
paths in the $p$-plane. Since we have assumed that there is only one resonance
pole at $p=p_R$, both paths in the $p$-plane start at the same point $p=p_0$
corresponding to $\eta=\eta_0$ and end at the same point $p=p_R$. The result
of the analytic continuation is, however, different along these paths, rendering
an unambiguous determination of the vertex function at $p=p_R$ impossible.      

The problem, which was discussed above, looks complicated but 
has a particularly simple solution. 
Let us go back to the last line of Eq.~(\ref{eq:two-terms}). 
It is immediately seen that the
ambiguity is caused by the expression $\eta\phi'(\eta)$, which is contained 
in the second term and depends on the
energy level index $n$. Moreover, the form of this expression is universal
(it does not depend on the interaction). Consequently, measuring the vertex
function for {\em two} different energy levels $n$ and $m$, and forming the
linear combination,
\eq
\bar V(p)=\frac{V_{nn}(p)\eta_m\phi'(\eta_m)-V_{mm}(p)\eta_n\phi'(\eta_n)}
 {\eta_m\phi'(\eta_m)-\eta_n\phi'(\eta_n)}\, ,
\en
one may immediately ensure that the culprit disappears. Namely,
$\bar V(p)$ is a low-energy polynomial in the variable $p^2$ up to a factor
$p^{-2}$, it does not depend on
the energy level (up to exponentially suppressed contributions), 
and its analytic continuation $p\to p_R$ 
into the complex $p$-plane
yields the infinite-volume vertex  function. To conclude, the problem with
the analytic continuation was circumvented by measuring the matrix elements
for two different energy levels.

Finally, we would like to note that the problem is milder in the case of 1+1
dimensions, see Ref.~\cite{1+1}. First, there are no finite fixed points
and no ambiguity emerges. Second, 
in Ref.~\cite{1+1} it has been shown that the
problematic contributions in 1+1 dimensions can be fit by a polynomial in $p$ (not $p^2$)
with $n$-dependent coefficients,
so the analytic continuation still can be performed (although it is a  more
subtle affair now, see Ref.~\cite{1+1} for the details).
No similar statement exists in the case of 3+1
dimensions. The subtraction trick can be used in 1+1 dimensions as well,
making the fit more straightforward (at the cost of measuring two energy levels instead of
one).

\section{Matrix elements at nonzero momentum transfer}
\label{sec:subtract}

We finally turn to the resonance matrix elements for non-zero momentum
transfer. It is convenient to work in the Breit frame ${\bf P}=-{\bf Q}$.
The vertex function in the infinite volume using dimensional regularization
 is given by (we again neglect the
numerators which do not affect the analytic properties):
\eq
I&=&\int\frac{d^d{\bf l}}{(2\pi)^d}\,
\frac{1}{8w({\bf l})w({\bf P}-{\bf l})w({\bf P}+{\bf l})}\,
\frac{1}{ (w({\bf P}-{\bf l})+w({\bf l})-P_0)
(w({\bf P}+{\bf l})+w({\bf l})-P_0)}\, ,
\nonumber\\[2mm]
p^*&=&\sqrt{\frac{s}{4}-m^2}\, ,\quad\quad s=P_0^2-{\bf P}^2\, ,
\en
where the $P_0\to P_0+i0$ prescription is implicit.
The finite-volume counterpart of this expression contains a sum over
the discrete momenta ${\bf l}$ instead of an integral. We note here once more
that a particular prescription is used to calculate this integral: the
integrand is first expanded in powers of the momenta, integrated over ${\bf l}$
and the resulting series is summed up again. Using this prescription, one may
present the above integral in the following form (consult, e.g.,
Ref.~\cite{cuspbig} for the technical details of similar calculations):
\eq
I&\!=\!&I_1+I_2\, ,
\nonumber\\[2mm]
I_1&\!=\!&\frac{1}{2P_0}\int\frac{d^d{\bf l}}{(2\pi)^d}\,
\frac{1}{4{\bf Pl}}\biggl\{\frac{1}{({\bf l}-\frac{\bf P}{2})^2
-\dfrac{1}{P_0^2}\,({\bf P}({\bf l}-\frac{\bf P}{2}))^2-(p^*)^2}
\nonumber\\[2mm]
&\!-\!&\frac{1}{({\bf l}+\frac{\bf P}{2})^2
-\dfrac{1}{P_0^2}\,({\bf P}({\bf l}+\frac{\bf P}{2}))^2-(p^*)^2}\biggr\}
\, ,
\nonumber\\[2mm]
I_2&\!=\!&-\frac{1}{2P_0}\int\frac{d^d{\bf l}}{(2\pi)^d}\,
\biggl\{\frac{1}{2w({\bf P}+{\bf l})(w({\bf P}-{\bf l})+w({\bf P}+{\bf l}))}\,
\frac{1}{({\bf l}-\frac{\bf P}{2})^2
-\dfrac{1}{P_0^2}\,({\bf P}({\bf l}-\frac{\bf P}{2}))^2-(p^*)^2}
\nonumber\\[2mm]
&\!+\!&\frac{1}{2w({\bf P}-{\bf l})(w({\bf P}-{\bf l})+w({\bf P}+{\bf l}))}\,
\frac{1}{({\bf l}+\frac{\bf P}{2})^2
-\dfrac{1}{P_0^2}\,({\bf P}({\bf l}+\frac{\bf P}{2}))^2-(p^*)^2}\biggr\}
\, .
\en
Explicit calculations yield the following result (on the second sheet):
\eq
I_1&=&\frac{1}{16\pi P_0|{\bf P}|}\,
\arctan\frac{-i\sqrt{s}|{\bf  P}|}{2P_0p^*}\, ,
\nonumber\\[2mm]
I_2&=&\frac{ip^*}{32\pi\sqrt{s}}\,\int_{-1}^{+1}dy\biggl\{
\frac{1}{w_2'(w_1'+w_2')}+\frac{1}{w_1''(w_1''+w_2'')}\biggr\}\, ,
\en
where
\eq
w_{1,2}'&=&\biggl(m^2+(p^*)^2\biggl(1+\frac{{\bf P}^2y^2}{s}\biggr)
\mp\frac{2P_0|{\bf P}|p^*y}{\sqrt{s}}\,\biggl(1\mp \frac{1}{2}\biggr)
+\biggl(1\mp \frac{1}{2}\biggr)^2{\bf P}^2\biggr)^{1/2}\, ,
\nonumber\\[2mm]
w_{1,2}''&=&\biggl(m^2+(p^*)^2\biggl(1+\frac{{\bf P}^2y^2}{s}\biggr)
\mp\frac{2P_0|{\bf P}|p^*y}{\sqrt{s}}\,\biggl(1\pm \frac{1}{2}\biggr)
+\biggl(1\pm \frac{1}{2}\biggr)^2{\bf P}^2\biggr)^{1/2}\, .
\en
Now let us consider the same quantities in a finite volume:
\eq
I_1^L&\!=\!&\frac{1}{2P_0}\frac{1}{L^3}\sum_{\bf l}
\frac{1}{4{\bf Pl}}\biggl\{\frac{1}{({\bf l}-\frac{\bf P}{2})^2
-\dfrac{1}{P_0^2}\,({\bf P}({\bf l}-\frac{\bf P}{2}))^2-(p^*)^2}
\nonumber\\[2mm]
&\!-\!&\frac{1}{({\bf l}+\frac{\bf P}{2})^2
-\dfrac{1}{P_0^2}\,({\bf P}({\bf l}+\frac{\bf P}{2}))^2-(p^*)^2}\biggr\}
\, ,
\nonumber\\[2mm]
I_2^L&\!=\!&-\frac{1}{2P_0}\frac{1}{L^3}\sum_{\bf l}
\biggl\{\frac{1}{2w({\bf P}+{\bf l})(w({\bf P}-{\bf l})+w({\bf P}+{\bf l}))}\,
\frac{1}{({\bf l}-\frac{\bf P}{2})^2
-\dfrac{1}{P_0^2}\,({\bf P}({\bf l}-\frac{\bf P}{2}))^2-(p^*)^2}
\nonumber\\[2mm]
&\!+\!&\frac{1}{2w({\bf P}-{\bf l})(w({\bf P}-{\bf l})+w({\bf P}+{\bf l}))}\,
\frac{1}{({\bf l}+\frac{\bf P}{2})^2
-\dfrac{1}{P_0^2}\,({\bf P}({\bf l}+\frac{\bf P}{2}))^2-(p^*)^2}\biggr\}
\, .
\en
Neglecting partial-wave mixing in the finite volume, the quantity $I_2^L$
can be rewritten as
\eq
I_2^L=\frac{1}{32\pi\sqrt{s}}\,\int_{-1}^{+1}dy\biggl\{
\frac{p^*\cot\phi^{\bf d}(s)}{w_2'(w_1'+w_2')}
+\frac{p^*\cot\phi^{-{\bf d}}(s)}{w_1''(w_1''+w_2'')}\biggr\}\, .
\en
Using the Gottlieb-Rummukainen equation, it is straightforward to ensure that
$I_2^L$ is a low-energy polynomial and its analytic continuation to $p=p_R$
gives the infinite-volume result $I_2$. On the contrary, $I_1^L$ does not have
the same property. For this term, we use the following trick. 
We define:
\eq
I_1^L=I_1^S+(I_1^L-I_1^S)=I_1^S+g\, ,\quad\quad
I_1^S=\frac{1}{16\pi P_0(p^*)^2}\,\biggl(\frac{p_R}{|{\bf P}|}\,
\arctan\frac{-i\sqrt{s}|{\bf P}|}{2P_0p_R}\biggr)p^*\cot\delta(s)\, .
\en
The quantity $I_1^S$ is a low-energy polynomial (up to a trivial overall factor
$(p^*)^{-2}$), and its analytic continuation to the pole on the second sheet
gives $I_1$, which is the value of the integral in the infinite
volume. Further, the quantity $g$ is dependent on the energy level, and is
universal (all derivative interactions factor out). Consequently, measuring
the vertex function for two different energy levels $n$ and $m$ in the Breit
frame, and forming the linear combination
\eq\label{eq:barV}
\bar V(p^*)=\frac{V_{nn}(p^*)g_m(p^*)-V_{mm}(p^*)g_n(p^*)}
{g_m(p^*)-g_n(p^*)}\, ,
\en
one sees that the culprit cancels out: $\bar V(p^*)$ is a polynomial up to a
factor $(p^*)^{-2}$, and its analytic continuation to the resonance pole
yields the vertex function in the infinite volume.

\section{Conclusions}
\label{sec:concl}

\begin{itemize}

\item[i)]
In this paper, by using the technique of the non-relativistic effective
Lagrangians in a finite volume, we were able to formulate a procedure for
extracting the resonance matrix elements on the lattice. The derivation was
restricted to the case of isolated resonances, lying in the region of the
applicability of the effective-range expansion.

\item[ii)]
As a demonstration of the usefulness of the non-relativistic EFT approach, we
have re-derived the L\"uscher equation in the moving frame (Gottlieb-Rummukainen
equation), as well as the relation of the time-like form factor to the matrix
elements measured on a Euclidean lattice.

\item[iii)]
A resonance pole is extracted in the following manner: by performing the
measurement of the energy levels at different volumes, and using L\"uscher's
formula, one extracts the function $p\cot\delta(s)$ at different values of
$p$. In the region of applicability of the effective-range expansion, 
which we have assumed here,
this function is a polynomial in the variable $p^2$: $p\cot\delta(s)=
A_0+A_1p^2+\cdots$ (for simplicity, we consider the S-wave).
The fit to the lattice data determines the coefficients $A_0,A_1,\cdots$.
The resonance pole position is then determined from the equation
\eq
p_R\cot\delta(s_R)=A_0+A_1p_R^2+\cdots=-ip_R\, .
\en
Note that a shortcut version of this procedure is to determine the zero of the
function $p\cot\delta(s)$ and to relate the width of a resonance to the derivative of
this function. At present, this shortcut version is routinely used to study
the resonance properties on the lattice. For narrow resonances, both
procedures give the same result.  

\item[iv)]
The case of the resonance form factors is more subtle. It has been demonstrated
that a straightforward
 analytic continuation of the matrix elements of the current between
the eigenstates of the Hamiltonian in a finite volume does not allow one to
determine resonance matrix elements unambiguously in 3+1 dimensions, and the
infinite volume limit can not be performed.

\item[v)]
The way to circumvent the above problem is to measure the matrix elements for
{\em two} (or, eventually, more) eigenstates. The extraction of the matrix
element proceeds in several steps:
\begin{itemize}

\item
Use the Breit frame, then extract matrix elements between at least two different
eigenstates, labeled by $n,m$,
 by using Eq.~(\ref{eq:E0}) (or its counterpart for excited
states).

\item 
Using Eq.~(\ref{eq:matrix-finite}), extract the quantities
$V_{nn}(p^*),V_{mm}(p^*)$ with $p^*=\bigl((E^2-{\bf
      P}^2)/4-m^2\bigr)^{1/2}$,
and $E=E_n~\mbox{or}~E_m$. Note that, in the Breit frame,
$V_{nn}(p^*),V_{mm}(p^*)$ depend only on $p^*$, as ${\bf P}$ is fixed.

\item
Form the linear combination $\bar V(p^*)$, using Eq.~(\ref{eq:barV}). Fit the
results of the measurements for different values of $L$ by using the formula
\eq
\bar V(p^*)=\frac{D_{-1}}{(p^*)^2}+D_0+D_1(p^*)^2+\cdots\, .
\en

\item
Calculate $\bar V^\infty=\bar V(p_R)$ by simply substituting $p^*=p_R$ in the
above expression.

\item
Finally, calculate the resonance form factor in the infinite volume by using
Eq.~(\ref{eq:matrix-infinite}).  

\item[vi)]
The procedure described above demands that the matrix elements between the
eigenstates are measured on the lattice at several different volumes and {\em
  at least for two different eigenstates}. We realize that, at present, this
requirement is rather challenging. However, in our opinion, it is still
important to have a clearly defined and mathematically rigorous procedure,
which will allow for a clean extraction of resonance form factors in the
future. Turning the argument around, our discussions demonstrate that the existing
lattice results for the resonance matrix elements should be put under 
renewed scrutiny.

\item[vii)]
It would be interesting to extend the discussion to the case of twisted
boundary conditions, which have proved advantageous in the calculations of 
form factors. Non-relativistic EFT is ideally suited for this purpose. We plan
to investigate this issue in the future.

\item[viii)]
In this paper, one has assumed that the effective-range expansion is valid for
the energies where the resonance is located. It would be interesting to extend
the range of applicability of the approach, by using e.g. conformal
mapping.

\end{itemize}

\bigskip

\section*{Acknowledgments}
The authors that  J. Gasser, 
M. G\"ockeler, Ch. Lang, H. Meyer, J. Pelaez, A. Sch\"afer and 
G. Schierholz for interesting discussions.
This work is partly supported by the EU
Integrated Infrastructure Initiative HadronPhysics3 Project  under Grant
Agreement no. 283286. We also acknowledge the support by DFG (SFB/TR 16,
``Subnuclear Structure of Matter'')
and by COSY FFE under contract 41821485 (COSY 106).
A.R. acknowledges  support of the Shota Rustaveli National Science Foundation 
(Project DI/13/6-100/11).

\end{itemize}


\begin{thebibliography}{99}

\bibitem{Gurtler:2008zz}
  M.~Gurtler {\it et al.}  [QCDSF Collaboration],
  PoS {\bf LATTICE2008} (2008) 051.

\bibitem{Alexandrou:2011ga}

  C.~Alexandrou, G.~Koutsou, H.~Neff, J.~W.~Negele, W.~Schroers and A.~Tsapalis,
  Phys.\ Rev.\  D {\bf 77} (2008) 085012
  [arXiv:0710.4621 [hep-lat]];

  C.~Alexandrou {\it et al.},
  Phys.\ Rev.\  D {\bf 79} (2009) 014507
  [arXiv:0810.3976 [hep-lat]];

  C.~Alexandrou,
  arXiv:1108.4112 [hep-lat].


\bibitem{Alexandrou:2010tj}

  C.~Alexandrou, G.~Koutsou, T.~Leontiou, J.~W.~Negele and A.~Tsapalis,
  Phys.\ Rev.\  D {\bf 76} (2007) 094511
  [Erratum-ibid.\  D {\bf 80} (2009) 099901]
  [arXiv:0706.3011 [hep-lat]];

  C.~Alexandrou, E.~B.~Gregory, T.~Korzec, G.~Koutsou, J.~Negele, T.~Sato and A.~Tsapalis,
  PoS {\bf LATTICE2010} (2010) 141
  [arXiv:1011.0411 [hep-lat]];


  C.~Alexandrou, E.~B.~Gregory, T.~Korzec, G.~Koutsou, J.~W.~Negele, T.~Sato and A.~Tsapalis,
  Phys.\ Rev.\ Lett.\  {\bf 107} (2011) 141601
  [arXiv:1106.6000 [hep-lat]].

\bibitem{em-Roper}
  H.~W.~Lin and S.~D.~Cohen,
  arXiv:1108.2528 [hep-lat].
\bibitem{Mandelstam}
  S.~Mandelstam,
  Proc.\ Roy.\ Soc.\ Lond.\  A {\bf 233} (1955) 248.


\bibitem{HuangWeldon}
  K.~Huang and H.~A.~Weldon,
  Phys.\ Rev.\  D {\bf 11} (1975) 257.


\bibitem{Luescher-torus}
  M.~L\"uscher,
  Nucl.\ Phys.\  B {\bf 354} (1991) 531.


\bibitem{He}
  C.~Liu, X.~Feng and S.~He,
  JHEP\  {\bf 0507} (2005) 011
  [arXiv:hep-lat/0504019];
  Int.\ J.\ Mod.\ Phys.\  A {\bf 21} (2006) 847
  [arXiv:hep-lat/0508022].

\bibitem{lage-KN}
  M.~Lage, U.-G.~Mei{\ss}ner and A.~Rusetsky,
  Phys.\ Lett.\  B {\bf 681}, 439 (2009)
  [arXiv:0905.0069 [hep-lat]].

\bibitem{lage-scalar}
  V.~Bernard, M.~Lage, U.-G.~Mei{\ss}ner and A.~Rusetsky,
  JHEP {\bf 1101} (2011) 019\\{}
  [arXiv:1010.6018 [hep-lat]].

\bibitem{oset}
  M.~D\"oring, U.-G.~Mei{\ss}ner, E.~Oset and A.~Rusetsky,
  Eur.\ Phys.\ J.\ A {\bf 47} (2011) 139
  [arXiv:1107.3988 [hep-lat]].

\bibitem{oset-others}
  A.~M.~Torres, L.~R.~Dai, C.~Koren, D.~Jido and E.~Oset,
  Phys.\ Rev.\ D {\bf 85} (2012) 014027
  [arXiv:1109.0396 [hep-lat]].

\bibitem{kappa}
  M.~D\"oring and U.-G.~Mei{\ss}ner,
  JHEP {\bf 1201} (2012) 009
  [arXiv:1111.0616 [hep-lat]].

\bibitem{Polejaeva-2}
  K.~Polejaeva and A.~Rusetsky,
  arXiv:1203.1241 [hep-lat], Eur. Phys. J. {\bf A} (2012), in print.

\bibitem{Michael}
  C.~Michael,
  Nucl.\ Phys.\ B {\bf 327} (1989) 515.

\bibitem{polejaeva}
  U.-G.~Mei{\ss}ner, K.~Polejaeva and A.~Rusetsky,
  Nucl.\ Phys.\  B {\bf 846} (2011) 1
  [arXiv:1007.0860 [hep-lat]].

\bibitem{entropy}
M.~Asakawa, T.~Hatsuda and Y.~Nakahara,
[arXiv:hep-lat/0011040v2];

  S.~Sasaki, K.~Sasaki, T.~Hatsuda and M.~Asakawa,
  Nucl.\ Phys.\ Proc.\ Suppl.\  {\bf 119} (2003) 302
  [arXiv:hep-lat/0209059];

  K.~Sasaki, S.~Sasaki and T.~Hatsuda,
  Phys.\ Lett.\  B {\bf 623} (2005) 208
  [arXiv:hep-lat/0504020].

\bibitem{Peardon}
  P.~Giudice, D.~McManus and M.~Peardon,
  arXiv:1204.2745 [hep-lat].

\bibitem{Lellouch}
  L.~Lellouch and M.~L\"uscher,
  Commun.\ Math.\ Phys.\  {\bf 219} (2001) 31
  [arXiv:hep-lat/0003023].

\bibitem{Sachrajda}
  C.~h.~Kim, C.~T.~Sachrajda, S.~R.~Sharpe,
  Nucl.\ Phys.\  {\bf B727} (2005) 218 (2005),  [hep-lat/0507006].

\bibitem{Meyer}
  H.~B.~Meyer,
  Phys.\ Rev.\ Lett.\  {\bf 107} (2011) 072002
  [arXiv:1105.1892 [hep-lat]].

\bibitem{Sharpe}
  M.~T.~Hansen and S.~R.~Sharpe,
  arXiv:1204.0826 [hep-lat].



\bibitem{1+1}
  D.~Hoja, U.-G.~Mei{\ss}ner and A.~Rusetsky,
  JHEP {\bf 1004} (2010) 050
  [arXiv:1001.1641 [hep-lat]].

\bibitem{gottlieb}
  K.~Rummukainen and S.~A.~Gottlieb,
  Nucl.\ Phys.\  B {\bf 450} (1995) 397
  [arXiv:hep-lat/9503028].


\bibitem{Fu}
  Z.~Fu,
  Phys.\ Rev.\ D {\bf 85} (2012) 014506
  [arXiv:1110.0319 [hep-lat]].

\bibitem{Prelovsek}
  L.~Leskovec and S.~Prelovsek,
  arXiv:1202.2145 [hep-lat].


\bibitem{Davoudi} 
Z.~Davoudi and M.~J.~Savage,
  Phys.\ Rev.\ D {\bf 84} (2011) 114502
  [arXiv:1108.5371 [hep-lat]].

\bibitem{Bob}
M. G\"ockeler, R. Horsley, M. Lage, U.-G. Mei{\ss}ner, P.E.L. Rakow,
      A. Rusetsky, G. Schierholz and J.M. Zanotti, in preparation.


\bibitem{Beane}
  S.~R.~Beane, P.~F.~Bedaque, A.~Parreno and M.~J.~Savage,
  Nucl.\ Phys.\  A {\bf 747} (2005) 55
  [arXiv:nucl-th/0311027].


\bibitem{lage-distributions}
  V.~Bernard, M.~Lage, U.-G.~Mei{\ss}ner and A.~Rusetsky,
  JHEP {\bf 0808} (2008) 024\\{}
  [arXiv:0806.4495 [hep-lat]].


\bibitem{cusp1}
  G.~Colangelo, J.~Gasser, B.~Kubis and A.~Rusetsky,
  Phys.\ Lett.\  B {\bf 638} (2006) 187
  [arXiv:hep-ph/0604084].


\bibitem{cuspbig}
  J.~Gasser, B.~Kubis and A.~Rusetsky,
  Nucl.\ Phys.\  B {\bf 850} (2011) 96
  [arXiv:1103.4273 [hep-ph]].

\bibitem{Bernard:1992qa}
  V.~Bernard, N.~Kaiser, J.~Kambor and U.-G.~Mei\ss ner,
  Nucl.\ Phys.\ B {\bf 388} (1992) 315.

\bibitem{luescher-2.}
  M.~L\"uscher,
  Commun.\ Math.\ Phys.\  {\bf 105} (1986) 153 (1986).

 
 
\bibitem{Oset-movingframe}
M. D\"oring, U.-G. Mei{\ss}ner, E. Oset and A. Rusetsky, in preparation.


\bibitem{Savage-formfactor}
  W.~Detmold and M.~J.~Savage,
  Nucl.\ Phys.\  A {\bf 743} (2004) 170
  [arXiv:hep-lat/0403005].




\bibitem{deDivitiis:2004rf}
  G.~M.~de Divitiis and N.~Tantalo,
  arXiv:hep-lat/0409154.

\bibitem{Christ:2005gi}
  N.~H.~Christ, C.~Kim and T.~Yamazaki,
  Phys.\ Rev.\  D {\bf 72} (2005) 114506
  [arXiv:hep-lat/0507009].


\end{thebibliography}
\end{document}